# Barium and europium abundances from the *Gaia*-ESO Survey using *The Cannon*

P. B. Das[1,2,3,⋆], F. Primas[1], G. De Silva[2,3,4], G. Guiglion[5,6,7], D. B. Zucker[2,3], and M. Ness[4]

[1] European Southern Observatory, Karl-Schwarzschild-Str. 2, 85748 Garching bei München, Germany
[2] School of Mathematical and Physical Sciences, Macquarie University, Sydney, NSW 2109, Australia
[3] Astrophysics and Space Technologies Research Centre, Macquarie University, Sydney, NSW 2109, Australia
[4] Research School of Astronomy and Astrophysics, The Australian National University, Canberra, ACT 2611, Australia
[5] Zentrum für Astronomie der Universität Heidelberg, Landessternwarte, Königstuhl 12, 69117 Heidelberg, Germany
[6] Max-Planck-Institut für Astronomie, Königstuhl 17, 69117 Heidelberg, Germany
[7] Leibniz-Institut für Astrophysik Potsdam (AIP), An der Sternwarte 16, 14482 Potsdam, Germany

Received 27 March 2026 / Accepted 16 June 2026

**ABSTRACT**

*Context.* Neutron-capture element abundances encode the time-dependent interplay between slow (s-) and rapid (r-) nucleosynthetic processes and provide critical constraints on Galactic chemical evolution. In particular, barium and europium provide important tracers of s-process and r-process nucleosynthetic enrichment, respectively. Large spectroscopic surveys enable robust statistical studies of these elements; however, determining reliable elemental abundances at intermediate spectral resolution remains challenging.

*Aims.* Our aim was to derive homogeneous and statistically well-characterised stellar parameters and chemical abundances of Ba and Eu (hereafter collectively referred to as stellar labels) for the *Gaia*-ESO Survey HR15N dataset.

*Methods.* We applied *The Cannon* to 36 263 FLAMES/GIRAFFE HR15N spectra ($R \approx 19\,200$, $\lambda \sim 6440$–6815 Å), using a model trained on 782 stars with high-quality reference stellar labels drawn from the *Gaia*-ESO DR5 homogenised catalogue. The model infers $T_{\rm eff}$, $\log g$, [Fe/H], $v \sin i$, $A$(Ba), and $A$(Eu); its reliability was quantified through multi-dimensional quality flags for the labels and detection strength of the Ba and Eu features. The uncertainties in the inferred stellar labels were obtained by combining the model's internal label-covariance errors with signal-to-noise-dependent dispersions.

*Results.* Our final catalogue contains stellar parameters for 36 263 spectra, with ~21 000 and ~400 reliable Ba and Eu abundances, respectively, satisfying multiple flagging criteria. We obtained median uncertainties of ~38 K in $T_{\rm eff}$, ~0.08 dex in $\log g$, ~0.04 dex in [Fe/H], ~0.06 dex in $A$(Ba), and ~0.02 dex in $A$(Eu). External validation against the *Gaia* Benchmark Stars, GALAH DR4, and multiple open clusters shows good agreement across independent reference datasets.

*Conclusions.* This study provides a comprehensive dataset of neutron-capture abundances in the Galactic disc observed with the *Gaia*-ESO HR15N set-up. The resulting catalogue includes newly determined Ba and Eu abundances that complement those available from previous analyses, thereby enabling well-characterised investigations of neutron-capture enrichment patterns in open clusters. The model framework can readily be adapted to upcoming wide-field spectroscopic surveys such as 4MOST, WEAVE, and WST.

## 1. Introduction

The chemical composition of long-lived stars preserves the nucleosynthetic signatures of the gas from which they formed, thereby providing a unique record of the formation and evolutionary history of the Milky Way. In the framework of Galactic archaeology, stellar abundance patterns act as fossil tracers of star formation, gas accretion, radial migration, and enrichment timescales across Galactic components (e.g. Freeman & Bland-Hawthorn 2002; Matteucci 2012; Rix & Bovy 2013; Bland-Hawthorn & Gerhard 2016). In particular, neutron-capture elements offer unique insights into time-delayed nucleosynthetic channels and the relative contributions of distinct astrophysical production sites.

The origin of neutron-capture elements is traditionally described within the framework established by Burbidge et al. (1957) and Cameron (1957), who identified the slow (s-) and rapid (r-) neutron-capture processes as the dominant formation pathways beyond the iron peak. The s-process operates at relatively low neutron densities and proceeds along the valley of $\beta$-stability, primarily occurring in asymptotic giant branch (AGB) stars (Busso et al. 1999; Karakas & Lattanzio 2014). In contrast, the r-process requires extreme neutron fluxes and is associated with highly energetic explosive environments. Recent theoretical and observational advances indicate that neutron-star mergers constitute an important astrophysical site of r-process nucleosynthesis (e.g. GW170817; Abbott et al. 2017; Cowan et al. 2021), although rare classes of core-collapse supernovae may also contribute (Thielemann et al. 2017).

Among neutron-capture species, barium and europium play key diagnostic roles. Ba is predominantly synthesised via the main s-process in low- and intermediate-mass AGB stars (Busso et al. 1999; Karakas & Lattanzio 2014), although additional early contributions from rotating massive stars may be important at low metallicity (Cescutti et al. 2013). Eu, in contrast, is predominantly produced via the r-process (Sneden et al. 2008;

⋆ Corresponding author: pbdrohan@gmail.com

Cowan et al. 2021). Candidate r-process sites include neutron-star mergers and rare classes of core-collapse supernovae, such as magneto-rotational or collapsar-like explosions. Neutron-star mergers are known to produce large amounts of r-process material per event; however, their generally longer delay times and low predicted event rates imply that their integrated contribution to Galactic chemical evolution remains uncertain (Cowan et al. 2021; Thielemann et al. 2017). Conversely, r-process production associated with rare classes of massive-star explosions can enrich the interstellar medium on shorter timescales, although their rates and yields remain uncertain. Therefore, the dominant Galactic r-process site remains an open question. Since these channels operate on different and potentially overlapping timescales, with AGB enrichment delayed relative to enrichment from massive-star explosions, the abundance ratios [Ba/Fe], [Eu/Fe], and especially [Eu/Ba] trace the combined effects of delayed s-process enrichment, prompt or delayed r-process production, and possible early Ba production from massive stars. These ratios thus provide constraints on the temporal build-up of neutron-capture elements in the Galaxy (Cescutti et al. 2006; Sneden et al. 2008).

In the Galactic halo, metal-poor stars exhibit large star-to-star scatter in r-process elements, indicating inhomogeneous early enrichment (Sneden et al. 2008; Hansen et al. 2018). In the disc, [Eu/Fe] decreases with increasing metallicity due to delayed Fe production from Type Ia supernovae (e.g. Battistini & Bensby 2016). Conversely, Ba exhibits more complex behaviour with notable enhancements in young open clusters (D'Orazi et al. 2009; Yong et al. 2012; Jacobson & Friel 2013; Mishenina et al. 2016; Spina et al. 2021). These unexpectedly high [Ba/Fe] values in young populations (e.g. Baratella et al. 2021; Lian et al. 2023; Chen et al. 2025) remain challenging to reconcile with standard chemical evolution models (Travaglio et al. 1999; Maiorca et al. 2012), thereby requiring analyses of non-local thermodynamic equilibrium (NLTE) effects, chromospheric activity, atomic data systematics, or modifications to AGB yields.

The advent of large spectroscopic surveys has significantly expanded our ability to investigate neutron-capture elements across Galactic populations. Surveys such as *Gaia*-ESO (Gilmore et al. 2012; Randich et al. 2013), GALAH (De Silva et al. 2015; Buder et al. 2018, 2021, 2025), RAVE (Steinmetz et al. 2020), SDSS/APOGEE (Majewski et al. 2017; Kollmeier et al. 2017; Abdurro'uf et al. 2022), LAMOST (Zhao et al. 2012), the *Gaia* mission (Gaia Collaboration 2016, 2023a,b), and *Gaia*-RVS (Recio-Blanco et al. 2016, 2023) have produced homogeneous stellar parameters and chemical abundances for millions of stars. These datasets enable statistically well-defined mapping of chemical gradients, age-abundance relations, and chemo-dynamical substructures across the Milky Way.

The *Gaia*-ESO Survey (GES) collected high-quality spectra of approximately 115 000 unique stars across the Milky Way (Gilmore et al. 2022; Randich et al. 2022) using the FLAMES facility at the ESO Very Large Telescope (Pasquini et al. 2002) by sampling major Galactic components including the halo, bulge, thin and thick discs, as well as multiple open clusters and globular clusters (Lardo et al. 2015; Bragaglia et al. 2022). The survey uses multiple GIRAFFE and UVES instrumental set-ups optimised for different stellar populations and science cases; in this work we focus on the FLAMES/GIRAFFE HR15N configuration (also known as H665.0), which has a resolving power of $R \approx 19\,200$ and covers approximately 6440–6815 Å. This wavelength range encompasses the Ba II at 6496.9 Å and Eu II at 6645.1 Å lines, as well as the Li I doublet at 6707.8 Å and H$\alpha$ at 6562.8 Å, making it particularly valuable for studies of young stars and neutron-capture element abundances. While UVES observations within GES provide high-fidelity abundance measurements (with $R \approx 47\,000$), the limited multiplexing of UVES (eight fibres per FLAMES plate) constrains target selection per field. By contrast, the HR15N configuration employs the GIRAFFE MEDUSA mode with up to 132 fibres simultaneously, enabling efficient observations of large cluster samples, which is critical for population-level chemical-evolution studies. However, deriving reliable abundances from HR15N spectra presents challenges; the intermediate spectral resolution can increase covariances between stellar parameters and abundances, while the limited number of diagnostic spectral lines constrains conventional spectroscopic abundance analyses.

Data-driven inference techniques provide a useful approach for addressing such limitations. Rather than performing a full physical spectral synthesis for each star, these approaches train a generative model directly on observed spectra using a reference sample of stars with known high-fidelity stellar labels[1], and then use the trained model to predict the labels for the larger sample of stellar spectra. *The Cannon* (Ness et al. 2015; Casey et al. 2016) is one such implementation of this methodology and has been widely applied to large surveys including APOGEE, LAMOST, *Gaia*–RVS, and GALAH (e.g. Ness et al. 2016; Ho et al. 2017a,b; Wheeler et al. 2020; Nandakumar et al. 2022; Das et al. 2025, 2026). By learning the flux–label mapping from a sample of high-quality stars, *The Cannon* can produce internally homogeneous stellar parameters and chemical abundances from intermediate-resolution and/or modest signal-to-noise ratio (S/N) data, provided sufficient spectral information is present.

*The Cannon* has been extended and applied in multiple studies that demonstrate its applicability to large spectroscopic datasets. For example, Casey et al. (2016) applied the framework to a sample of 87 563 red giant stars, inferring 17 stellar labels, including 15 individual elemental abundances with associated uncertainties, even for spectra with reduced S/N. Wheeler et al. (2020) employed *The Cannon* to determine chemical abundances for nearly 3.9 million LAMOST stars, covering five nucleosynthetic channels while using GALAH as the reference dataset. Nandakumar et al. (2022) utilised *The Cannon* to merge APOGEE and GALAH data into homogeneous catalogues that reconcile stellar parameters and abundances between the surveys. More recent analyses have further highlighted the capability of *The Cannon* for predicting multiple elemental abundances, including $\alpha$- and Fe-peak elements for roughly 800 000 *Gaia*-RVS spectra (Das et al. 2025), and for neutron-capture elements such as Nd, Ce, and Zr in a subset of approximately 350 000 *Gaia*-RVS spectra (Das et al. 2026). Previous data-driven studies specifically targeting GES data have employed alternative machine-learning approaches. These include Ambrosch et al. (2023), who applied convolutional neural networks (CNNs) to GIRAFFE spectra to derive homogeneous $T_{\rm eff}$, $\log g$, [Fe/H], [Mg/Fe], and [Al/Fe] estimates for 30 115 stellar spectra; Nepal et al. (2023), who trained a CNN on GES iDR6 HR15N spectra to infer $T_{\rm eff}$, $\log g$, [Fe/H], and A(Li) for ~40 000 stars; and Candebat et al. (2024), who developed a conditional invertible neural network (cINN) for GES GIRAFFE HR10 and HR21 spectra to simultaneously infer stellar parameters, elemental abundances, and their associated uncertainties.

In this work, we applied a quadratic model of *The Cannon* to the GES HR15N spectroscopic dataset to derive stellar parameters and elemental abundances. Our approach is anchored on a selected training set of HR15N spectra with high-quality reference labels from the GES DR5 homogenised catalogue (Hourihane et al. 2023), ensuring that the inferred labels remain on the established GES abundance scale. We employed statistical cross-validation and quantitative reliability metrics to flag unphysical extrapolations and characterise the precision and systematic uncertainties in the inferred labels. The resulting catalogue primarily includes stars from GES cluster observations, covering open clusters spanning ages from a few million to several billion years and galactocentric radii extending from the inner to the outer Galactic disc, thereby enabling investigations of chemical evolution trends across a broad range of cluster ages.

[1] The term 'labels' refers collectively to stellar parameters and elemental abundances used in *The Cannon* model.

This paper is structured as follows. Section 2 describes the details of the data used, the pre-processing steps applied to the HR15N spectra, and the construction of both the training and test datasets for our analysis. Section 3 details the implementation of *The Cannon*, including the model architecture and the training procedures. Section 4 discusses the various flagging conditions adopted for *The Cannon*-derived stellar labels in our final catalogue and their recommended applications. Section 5 explains the total uncertainties in *The Cannon*-derived stellar labels by incorporating the internal covariance errors and external S/N-dependent error profiles of the stellar labels. Section 6 analyses *The Cannon*-derived labels ($T_{\rm eff}$, $\log g$, $v \sin i$, and elemental abundances for [Fe/H], $A$(Ba), and $A$(Eu)) in the final catalogue. It also presents the validation of the model predictions against the *Gaia* Benchmark Stars, the GALAH DR4 catalogue, and multiple open clusters. Finally, Section 7 summarises our results and conclusions. The resulting homogeneous neutron-capture elemental abundance catalogue (Ba and Eu), including new *The Cannon*-derived Ba and Eu abundance measurements presented in this work, provides a comprehensive dataset for chemical-evolution studies and for testing data-driven spectroscopic methodologies at HR15N-like resolution. Given its spectral characteristics and model framework, the approach presented here is adaptable to other wide-field surveys such as 4MOST (de Jong et al. 2019), WEAVE (Dalton et al. 2018), and WST (Bacon et al. 2024), which operate at comparable resolving power and cover a wide range in wavelength.

## 2. Data

We adopted stellar parameters and elemental abundances from the GES DR5 catalogue[2],[3] (Hourihane et al. 2023) to build the selection criteria for our training sample. Within GES, multiple independent analysis nodes derived stellar parameters and abundances using diverse spectroscopic pipelines, adopting a common line list and coordinated analysis strategy. The resulting measurements were subsequently homogenised to place all results on a consistent reference scale (Smiljanic et al. 2014; Lanzafame et al. 2015; Gilmore et al. 2022; Randich et al. 2022; Worley et al. 2024). We made use of the recommended stellar parameters ($T_{\rm eff}$, $\log g$, $v \sin i$, and [Fe/H]) and the abundances[4] $A$(Ba) and $A$(Eu) from the catalogue.

Our initial sample consisted of 39 008 spectra[5] observed with the HR15N configuration, covering approximately 6440–6815 Å at a resolving power of $R \approx 19\,200$. The corresponding normalised spectra were used as the input fluxes for our analysis, while the GES catalogue provided the reference labels for training and validation. We excluded 2745 spectra from the sample due to missing radial velocities (`VRAD`), leaving a total of 36 263 spectra in the final working sample.

### 2.1. Pre-processing of HR15N spectra

All 36 263 HR15N spectra were corrected to the rest frame using the radial velocities (`VRAD`) and the corresponding offsets (`VRAD_OFFSET`) from the GES catalogue, ensuring a consistent wavelength scale. These spectra were trimmed to the wavelength range 6450–6750 Å and interpolated onto a uniform wavelength grid with a pixel spacing of 0.05 Å. This ensured that all spectra were placed on the same wavelength grid, which is necessary for training *The Cannon* model. The local continuum level of each spectrum was estimated as the median of the highest 20% of flux pixels and used to identify extreme flux values. Pixels lying more than five times their corresponding flux uncertainty ($\sigma$) above this level, together with pixels having negative flux values, were replaced with the estimated continuum level.

### 2.2. Training sample

From the sample of HR15N spectra, a high-fidelity training subset was selected by incorporating multiple quality conditions. Stars in GES flagged with problems relevant to spectral quality and/or estimated stellar parameters and abundances (incorporating `SFLAGS`; see Hourihane et al. 2023 for details) were not included in the training sample. Further screening conditions were implemented on the sample to ensure the reliability of both the reference stellar parameters and elemental abundances: stars were retained as part of the training sample if their GES parameters satisfied the following ranges: 3500 ≤ `TEFF` (K) ≤ 7000, 0.0 ≤ `LOGG` (dex) ≤ 5.0, `VSINI` (km s$^{-1}$) ≥ 0, –2.0 ≤ `FEH` (dex) ≤ 0.5, –0.5 ≤ `BA2` (dex) ≤ 3.0, and –0.5 ≤ `EU2` (dex) ≤ 3.0. Here, `TEFF`, `LOGG`, `VSINI`, `FEH`, `BA2`, and `EU2` refer to effective temperature ($T_{\rm eff}$), surface gravity ($\log g$), projected rotational velocity ($v \sin i$), metallicity ([Fe/H]), and absolute abundances of Ba II ($A$(Ba)) and Eu II ($A$(Eu)), respectively.

The corresponding uncertainties were additionally required to satisfy the conditions: `e_TEFF` ≤ 100 K, `e_LOGG` ≤ 0.3 dex, `e_FEH` ≤ 0.2 dex, `e_BA2` ≤ 0.3 dex, and `e_EU2` ≤ 0.3 dex. Stars with radial velocity uncertainties (`E_VRAD`) exceeding 0.5 km s$^{-1}$ or with spectra having `S/N` < 30 were also excluded. Finally, stars with projected rotational velocity satisfying `LIM_VSINI = 0` (which means the value is a detection and not a limit) were selected to ensure reliable values for `VSINI`.

After applying all these selection criteria, the final training sample comprised 782 stars and their corresponding stellar labels from the GES homogenised catalogue, which were used to develop *The Cannon* model.

### 2.3. Test sample

The test sample comprises the complete set of 36 263 HR15N spectra. As mentioned in Sect. 2.1, this sample excludes stars with missing radial velocity information, since *The Cannon* requires spectra to be corrected for radial velocity and placed on a common wavelength grid for uniform analysis.

[2] Based on data obtained from the ESO Science Archive Facility with DOI: `https://doi.org/10.18727/archive/25`

[3] `https://cdsarc.cds.unistra.fr/viz-bin/cat/J/A+A/676/A129`

[4] Absolute abundances are on the logarithmic scale $A(X) = \log_{10}(N_{\rm X}/N_{\rm H}) + 12$, where $N_{\rm X}$ and $N_{\rm H}$ are the number densities of element X and hydrogen, respectively.

[5] `http://ges.roe.ac.uk/index.html`; `http://casu.ast.cam.ac.uk/gaiaeso`

## 3. Application of *The Cannon* to HR15N spectra

*The Cannon* (Ness et al. 2015; Casey et al. 2016) is a data-driven framework that constructs a generative model of stellar spectra from a reference set of stars with well-determined labels. We note that *The Cannon* does not perform *ab initio* spectral synthesis, but instead learns an empirical flux-label mapping anchored to the reference abundance scale of the training set.

In this study, the training set consists of high-quality HR15N spectra (see Sect. 2.2) with input stellar labels: $T_{\rm eff}$, $\log g$, [Fe/H], $v \sin i$, $A$(Ba), and $A$(Eu). By accounting for observational uncertainties, the method enables reliable propagation of stellar labels from higher S/N training spectra to lower S/N test spectra, obtained with the same instrumental set-up, provided that the test spectra are adequately represented by the training sample. This enables homogeneous stellar-label inference across the full HR15N sample while retaining the GES reference scale. During training, the normalised flux at each wavelength pixel $\lambda$ for each star $n$ is modelled as a quadratic polynomial function of the stellar-label vector $\boldsymbol{l}_n$,

$$G_{n\lambda} = f(\boldsymbol{l}_n) \cdot \boldsymbol{\theta}_\lambda + \epsilon_{n\lambda}, \tag{1}$$

where $G_{n\lambda}$ denotes the observed flux, $f(\mathbf{l}_n)$ represents the vector of polynomial terms constructed from the stellar labels, and $\boldsymbol{\theta}_\lambda$ is the corresponding coefficient vector at wavelength pixel $\lambda$. The residual term $\epsilon_{n\lambda}$ accounts for deviations between the model and the observations. The total variance associated with each pixel combines the reported measurement uncertainty $\sigma_{n\lambda}$ with an additional term $s_\lambda$, which captures intrinsic scatter arising from astrophysical effects not explicitly described by the model:

$$\sigma^2_{\rm tot,n\lambda} = \sigma^2_{n\lambda} + s^2_\lambda, \qquad \epsilon_{n\lambda} \sim \mathcal{N}(0, \sigma^2_{\rm tot,n\lambda}). \tag{2}$$

The model parameters $\boldsymbol{\theta}_\lambda$ and $s_\lambda$ are determined independently at each wavelength pixel by maximising the likelihood over the training set:

$$\ln p\Big(\{G_{n\lambda}\} \,\Big|\, \boldsymbol{\theta}_\lambda, \{f(\boldsymbol{l}_n)\}, s^2_\lambda\Big) = -\frac{1}{2} \sum_{n=1}^{N_{\rm train}} \left[ \frac{(G_{n\lambda} - f(\boldsymbol{l}_n)\cdot\boldsymbol{\theta}_\lambda)^2}{\sigma^2_{n\lambda} + s^2_\lambda} + \ln\Big(\sigma^2_{n\lambda} + s^2_\lambda\Big) \right]. \tag{3}$$

Here, $N_{\rm train}$ denotes the number of spectra in the training sample. After training, the coefficients $\boldsymbol{\theta}_\lambda$ and scatter terms $s_\lambda$ are fixed. For a new spectrum $m$, the stellar-label vector $\boldsymbol{l}_m$ is obtained by minimising the weighted residuals between the observed and model fluxes over all wavelength pixels:

$$\boldsymbol{l}_m = \arg\min_{\boldsymbol{l}} \sum_{\lambda=1}^{N_\lambda} \frac{[G_{m\lambda} - f(\boldsymbol{l}) \cdot \boldsymbol{\theta}_\lambda]^2}{\sigma^2_{m\lambda} + s^2_\lambda}. \tag{4}$$

Here, $N_\lambda$ denotes the number of wavelength pixels included in the model fit.

In this work, we employed a second-order polynomial *The Cannon*[6] model (Ness et al. 2015; Casey et al. 2016), trained on 782 HR15N spectra (see Sect. 2.2) using six stellar labels: ($T_{\rm eff}$, $\log g$, $v \sin i$, [Fe/H], $A$(Ba), and $A$(Eu)). This resulted in a total of 28 coefficients per wavelength pixel: six linear coefficients (one per label), fifteen cross-term coefficients for all unique label pairs, six quadratic coefficients (one per label), and a constant term.

Prior to training the model, specific wavelength windows were also masked to ensure that only features reliably represented in the training set influenced the model without affecting the inferred stellar labels. This is particularly important given that the training sample spans a wide range of effective temperatures and surface gravities, where strong spectral features characteristic of some stars, such as very young or very hot/-cool stars, could potentially bias the inferred labels and introduce uncertainties. In particular, a ±3 Å window around the H$\alpha$ line (6562.8 ± 3 Å) and a narrow window around the Li I line at 6707.8 Å (6707–6709 Å) were masked in the model. The H$\alpha$ masking region was chosen to exclude the line core, where chromospheric emission and activity-related variability are most prominent at the HR15N resolving power ($R \approx 19\,200$). This preserves nearby spectral features that contain information on the stellar labels and avoids excessive masking of the surrounding continuum and metallic lines. Very young or strongly accreting stars may exhibit broader H$\alpha$ emission extending beyond this range. Such cases are not fully addressed by the masking alone and may produce poorer global spectral fits; the flagging criteria and the reduced-$\chi^2$ values provided in the catalogue offer additional diagnostics. The Li I region was masked to avoid biases from Li-enhanced stars, while retaining the surrounding continuum and neighbouring absorption features. The blue (6450–6460 Å) and the red edges (6740–6750 Å) of the wavelength range were masked to avoid any potential edge effects. After masking, all non-finite flux values were replaced with unity and the corresponding inverse variances were set to zero, ensuring that masked pixels do not contribute to the model fitting. This effectively removes regions prone to systematic variability or physical behaviour not encoded in the training labels, thereby improving the reliability of the inferred stellar parameters and elemental abundances.

To assess the internal consistency of the training sample, we performed a five-fold cross-validation (similar to Ho et al. 2017a; Buder et al. 2018; Nandakumar et al. 2022; Das et al. 2025, 2026) of *The Cannon* predictions for the training sample of 782 stars. In this procedure, the training sample was randomly divided into five groups. For each iteration, one group was left out as a test set, while the model was trained on the remaining four groups. The trained model was then applied to the omitted group to predict stellar labels. This process was repeated for all five independent groups. Finally, once all the iterations were completed, we obtained the estimates of the root-mean-square error (RMSE) and bias, which offer an assessment of the precision of the inferred stellar parameters and elemental abundances within the training sample.

## 4. Flagging of *The Cannon* labels

Once *The Cannon* was trained on the 782 HR15N spectra, the resulting model coefficients were used to predict the six stellar labels ($T_{\rm eff}$, $\log g$, $v \sin i$, [Fe/H], $A$(Ba), and $A$(Eu)) for the test sample of 36 263 HR15N spectra. This produced a catalogue of homogeneous stellar parameters and elemental abundances anchored to the high-fidelity GES reference scale.

The reliability of predicted labels, however, depends on their proximity to the training distribution in multi-dimensional label space. To quantify this, we define a pairwise distance metric $d$

[6] https://github.com/andycasey/AnniesLasso

**Table 1.** Calibrated distance thresholds for quality flags implemented in the model.

| Flag | $N_{\text{labels}}$ | Threshold ($D_{\text{thr}}$) |
|---|---|---|
| `flag_atm` | 4 | 9.70 |
| `flag_Ba` | 5 | 10.87 |
| `flag_Eu` | 5 | 10.38 |

**Notes.** Thresholds are set to twice the standard deviation of the training-sample local-distance distributions; each local distance is the mean of the ten smallest pairwise distances, with each label dimension normalised by its five-fold cross-validation RMSE.

(e.g. Ho et al. 2017a; Nandakumar et al. 2022; Das et al. 2025, 2026) between predicted *Cannon* labels ($\mathbf{l}_{\text{Cannon}}$) and training labels ($\mathbf{l}_{\text{train}}$), normalised by the root mean square error (RMSE) values obtained from the five-fold cross-validation,

$$d = \sum_i \frac{(l_{i,\text{Cannon}} - l_{i,\text{train}})^2}{\text{RMSE}_i^2}, \tag{5}$$

where $\text{RMSE}_i$ represents the characteristic cross-validation uncertainty for label $i$. For each test spectrum, $d$ is calculated relative to every spectrum in the training sample. The local distance ($D$) assigned to the spectrum is then defined as the mean of its ten smallest pairwise distances in the relevant label space.

We define two complementary distance-based flagging criteria. First, for the stellar parameter quality flag, we evaluate Eq. (5) using the four labels $T_{\text{eff}}$, $\log g$, $v \sin i$, and [Fe/H]. We denote the resulting pairwise distance by $d_{\text{atm}}$ and the corresponding local distance, defined as the mean of the ten smallest pairwise distances, by $D_{\text{atm}}$. The corresponding flag for the stellar parameters is then defined as:

$$\texttt{flag_atm} = \begin{cases} 0, & D_{\text{atm}} \le D_{\text{thr,atm}} \\ & \text{(stellar-parameter interpolation)} \\ 1, & D_{\text{atm}} > D_{\text{thr,atm}} \\ & \text{(stellar-parameter extrapolation)} \end{cases} \tag{6}$$

The threshold $D_{\text{thr,atm}}$ is calibrated from the training sample by calculating $D_{\text{atm}}$ for each training star and setting the threshold to twice the standard deviation of the resulting distribution of local distances.

Second, element-specific flags evaluate abundance reliability by including the element of interest ($X$):

$$d_{\text{X}} = d_{\text{atm}} + \frac{(A(X)_{\text{Cannon}} - A(X)_{\text{train}})^2}{\text{RMSE}_X^2}, \tag{7}$$

$$\texttt{flag_X} = \begin{cases} 0, & D_{\text{X}} \le D_{\text{thr,X}} \quad \text{(abundance interpolation)} \\ 1, & D_{\text{X}} > D_{\text{thr,X}} \quad \text{(abundance extrapolation)} \end{cases}. \tag{8}$$

Here $X$ = Ba or Eu. The local element-specific distance $D_{\text{X}}$ is defined as the mean of the ten smallest pairwise values of $d_{\text{X}}$ in the corresponding five-dimensional label space, following the same distance-averaging procedure used for $D_{\text{atm}}$. The element-specific thresholds $D_{\text{thr,X}}$ are calibrated analogously from their respective training-sample local-distance distributions. The calibrated thresholds are reported in Table 1.

To further assess whether the measured Ba II and Eu II line features correspond to genuine detections or should be treated as non-detections, we apply a per-pixel detection criterion. A line is considered detected if pixels within the line centre window (6496.9±0.20 Å for Ba II, and 6645.1±0.20 Å for Eu II) show a flux depth ≥3% below the normalised continuum and exceed three times the local flux uncertainty ($\sigma$). Lines failing either criterion are classified as non-detections. The final catalogue includes the columns `3-sigma_X` (for `X = Ba, Eu`), where `3-sigma_X = 0` indicates detections, and `3-sigma_X = 1` denotes non-detections. Figure 1 shows examples of stellar spectra with `3-sigma_X = 0/1` processed in our analysis for `X = Ba` and `X = Eu`.

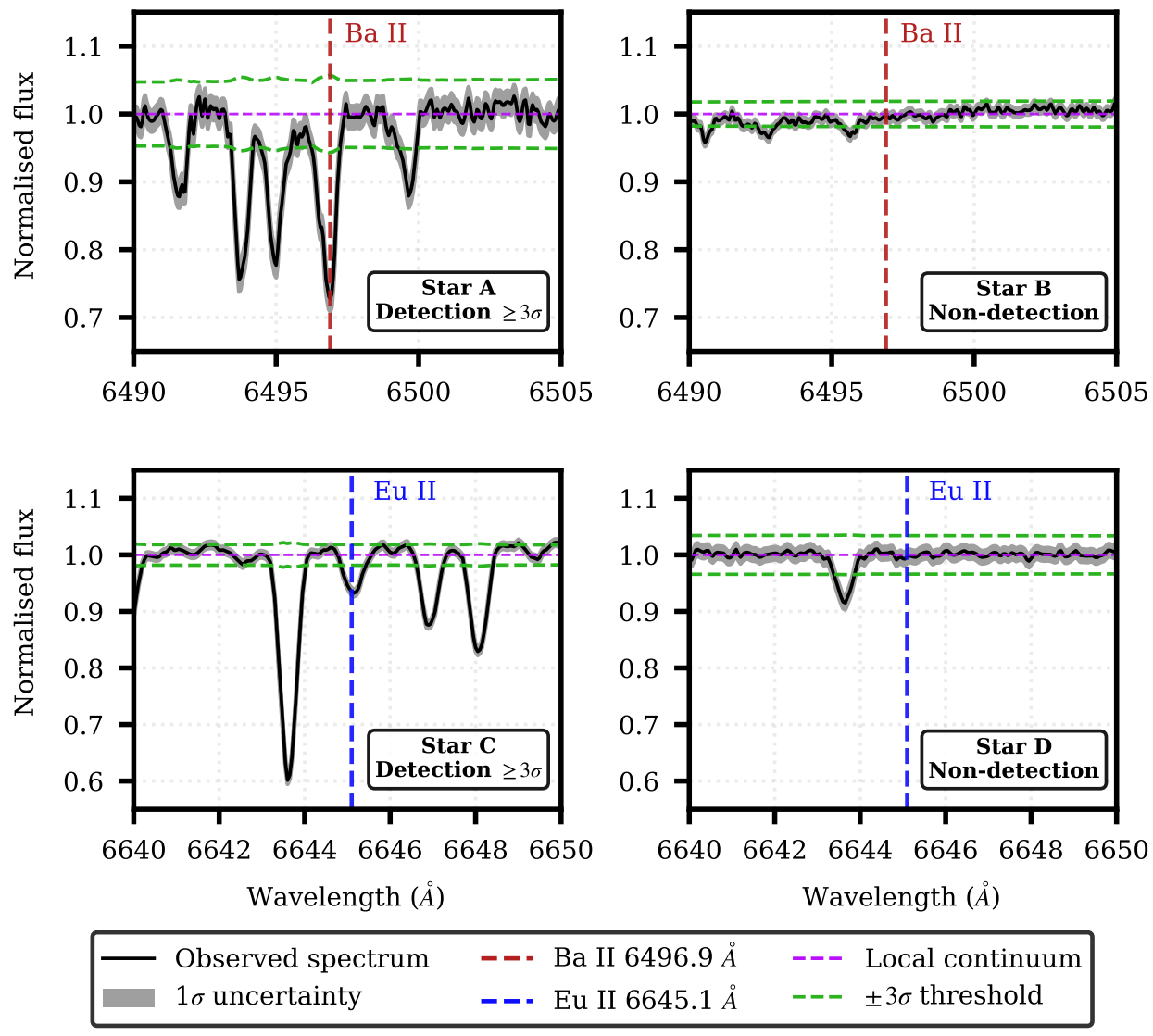

**Fig. 1.** Normalised HR15N spectra (solid black line). The grey shaded area indicates the 1$\sigma$ flux uncertainty around the Ba II line at 6496.9 Å (top panels) and the Eu II line at 6645.1 Å (bottom panels) for four representative stars. The panels for Star A and Star C show examples in which the Ba II and Eu II line depths exceed the 3$\sigma$ threshold, respectively, while the panels for Star B and Star D show examples in which the corresponding line depths remain below the 3$\sigma$ threshold. The vertical dashed lines mark the rest wavelengths of the Ba II and Eu II transitions (red and blue, respectively). The horizontal green dashed curves indicate the ±3$\sigma$ detection thresholds computed relative to the normalised continuum level. The four example stars correspond to Star A (ObjID: 18505731+1018278, $T_{\text{eff}}$ = 6426 K, $\log g$ = 4.02 dex, [Fe/H] = 0.03 dex), Star B (ObjID: 18051661−2419537, $T_{\text{eff}}$ = 6991 K, $\log g$ = 4.07 dex, [Fe/H] = −0.38 dex), Star C (ObjID: 10365793−5856584, $T_{\text{eff}}$ = 4417 K, $\log g$ = 1.99 dex, [Fe/H] = −0.21 dex), and Star D (ObjID: 11054909−5909492, $T_{\text{eff}}$ = 6532 K, $\log g$ = 4.22 dex, [Fe/H] = −0.16 dex).

We also provide the reduced $\chi^2$ values (`rchi_sq`) for the entire spectral wavelength range to quantify the statistical fit between the observed and *The Cannon*-generated spectra. In addition, to assess the quality of the fit within the element-specific wavelength regions, we compute the reduced $\chi^2$ values within the wavelength intervals 6496.9±0.20 Å for Ba II and 6645.1±0.20 Å for Eu II. These quantities are provided in the columns under `rchi_sq_X` for `X = Ba and Eu`.

Table A.1 gives a detailed description of all columns in our final catalogue. We recommend the following selection criteria for reliable measurements:

– Stellar parameters: `flag_atm = 0`
– Absolute abundances $A(X)$: `flag_X = 0`, `3-sigma_X = 0`
– Relative abundances [$X$/Fe]: `flag_atm = 0`, `flag_X = 0`, `3-sigma_X = 0`

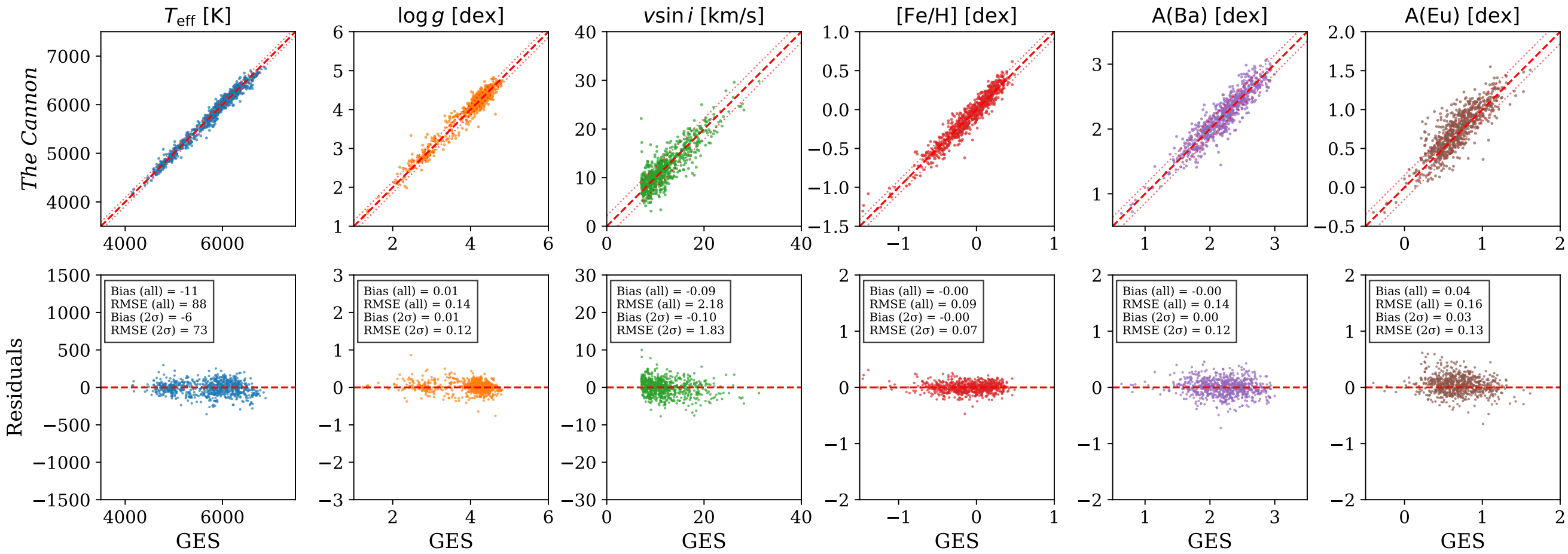

**Fig. 2.** One-to-one comparison of *The Cannon*-derived stellar labels vs *GES* reference labels for $T_{\rm eff}$, $\log g$, $v \sin i$, [Fe/H], $A$(Ba), and $A$(Eu). The top panels show *The Cannon*-derived and GES reference values for the stellar labels. The bottom panels display the residuals as a function of the reference labels. Bias and RMSE for both the full sample and the $2\sigma$ clipped stars are indicated, demonstrating reliable label recovery.

These quality flags enable robust sample selection for different scientific applications, while excluding abundance estimates that lie outside the label space sampled by the training set, as well as non-detections of the corresponding spectral features. In addition, we provide the label distances (`atm_dist`, `Ba_dist`, and `Eu_dist`) to enable further refinement of these thresholds depending on the desired level of quality control.

## 5. Error budget in the stellar labels

The per-pixel intrinsic scatter ($s_\lambda$) from *The Cannon* (see Sect. 3) quantifies model residuals but does not account for the complete S/N-dependent propagation into label uncertainties. To estimate this contribution, we performed noise resampling on the complete test sample of 36 263 spectra. We generated $N_{\rm rep} = 100$ realisations per spectrum by drawing the flux at each wavelength pixel $\lambda$ for each spectrum $n$ from a Gaussian distribution with mean $G_{n\lambda}$ and standard deviation $\sigma_{n\lambda}$. Each realisation was then processed with the trained *The Cannon* model to produce $N_{\rm rep}$ label estimates per spectrum. For spectrum $n$ and label $l$, the S/N-dependent uncertainty was obtained as the dispersion across these 100 realisations,

$$\sigma^{\rm obs}_{n,l} = \sqrt{\frac{1}{N_{\rm rep}-1}\sum_{k=1}^{N_{\rm rep}}\left(l^{(k)}_{n,l} - \overline{l_{n,l}}\right)^2}, \tag{9}$$

where $l^{(k)}_{n,l}$ is the label estimate obtained from the $k$th realisation and $\overline{l_{n,l}}$ is the mean of the $N_{\rm rep}$ label estimates. This was complemented by the internal model uncertainty $\sigma^{\rm int}_{n,l}$, extracted from *The Cannon* label-space covariance matrix. Finally, we calculated the total uncertainties by combining the internal model and S/N-dependent uncertainties in quadrature:

$$\sigma^{\rm tot}_{n,l} = \sqrt{\left(\sigma^{\rm obs}_{n,l}\right)^2 + \left(\sigma^{\rm int}_{n,l}\right)^2}. \tag{10}$$

Similar approaches to estimating S/N-dependent uncertainties have also been discussed in Ness et al. (2015), Wheeler et al. (2020), Manea et al. (2024), and Das et al. (2025, 2026). Table A.1 includes the derived total uncertainties for the stellar labels in the columns `e_Teff`, `e_logg`, `e_vsini`, `e_FeH`, `e_A(Ba)`, and `e_A(Eu)`, respectively. For spectra satisfying `flag_atm = 0`, we obtained median uncertainties of ~38 K in $T_{\rm eff}$, ~0.08 dex in $\log g$, and ~0.04 dex in [Fe/H]. For Ba and Eu measurements satisfying their respective flagging criteria, the median uncertainties were ~0.06 dex in $A$(Ba) and ~0.02 dex in $A$(Eu), respectively.

We also included relative abundance ratios ([Ba/Fe] and [Eu/Fe]) in our final catalogue using $A$(Ba), $A$(Eu), and [Fe/H]:

$$[{\rm Ba/Fe}] = A({\rm Ba}) - [{\rm Fe/H}] - 2.17, \tag{11}$$
$$[{\rm Eu/Fe}] = A({\rm Eu}) - [{\rm Fe/H}] - 0.52, \tag{12}$$

where $[{\rm X/Fe}] = A({\rm X}) - [{\rm Fe/H}] - A({\rm X})_\odot$ quantifies the enrichment of element X relative to iron in the star compared to the Sun, adopting solar abundances (Grevesse et al. 2007) of $A({\rm Ba})_\odot = 2.17$ dex and $A({\rm Eu})_\odot = 0.52$ dex (same as the values adopted in GES). The uncertainties in these abundance ratios are obtained by propagating the individual errors in quadrature:

$$\sigma_{[{\rm Ba/Fe}]} = \sqrt{\left(\sigma_{A({\rm Ba})}\right)^2 + \left(\sigma_{[{\rm Fe/H}]}\right)^2}, \tag{13}$$

$$\sigma_{[{\rm Eu/Fe}]} = \sqrt{\left(\sigma_{A({\rm Eu})}\right)^2 + \left(\sigma_{[{\rm Fe/H}]}\right)^2}, \tag{14}$$

where $\sigma_{A({\rm Ba})}$, $\sigma_{A({\rm Eu})}$, and $\sigma_{[{\rm Fe/H}]}$ are the total uncertainties for $A$(Ba), $A$(Eu), and [Fe/H], respectively, as described in Eq. (10). These propagated errors are included in the final catalogue in the columns `e_BaFe` and `e_EuFe`, respectively.

## 6. Analysis and validation of *The Cannon* labels

Figure 2 shows the one-to-one plot for *The Cannon* estimates of the six stellar labels ($T_{\rm eff}$, $\log g$, $v \sin i$, [Fe/H], $A$(Ba), and $A$(Eu)) against the GES reference input labels for the training sample of 782 stars. In each panel, we report the RMSE and bias value for the entire training sample as well as for the sub-sample within the $2\sigma$ distribution. We obtain low RMSE values for the six labels: $T_{\rm eff} \sim 73$ K, $\log g \sim 0.12$ dex, $v \sin i \sim 1.83$ km s$^{-1}$, [Fe/H] ~ 0.07 dex, $A$(Ba) ~ 0.12 dex, and $A$(Eu) ~ 0.13 dex for the $2\sigma$ sub-sample. Similarly, we also show the one-to-one plot between the derived abundance ratios [Ba/Fe], [Eu/Fe] and [Eu/Ba] from *The Cannon* and the GES in Fig. 3. These small cross-validation residuals support the internal consistency of the label predictions. We also show the Kiel diagrams for the training sample comparing *The Cannon*-derived labels against the reference GES labels ($T_{\rm eff}$, $\log g$, and [Fe/H]) in Fig. 4. The Kiel diagrams show similar overall trends in the label spaces for the stellar parameters between *The Cannon* and the GES, thereby further validating the training sample.

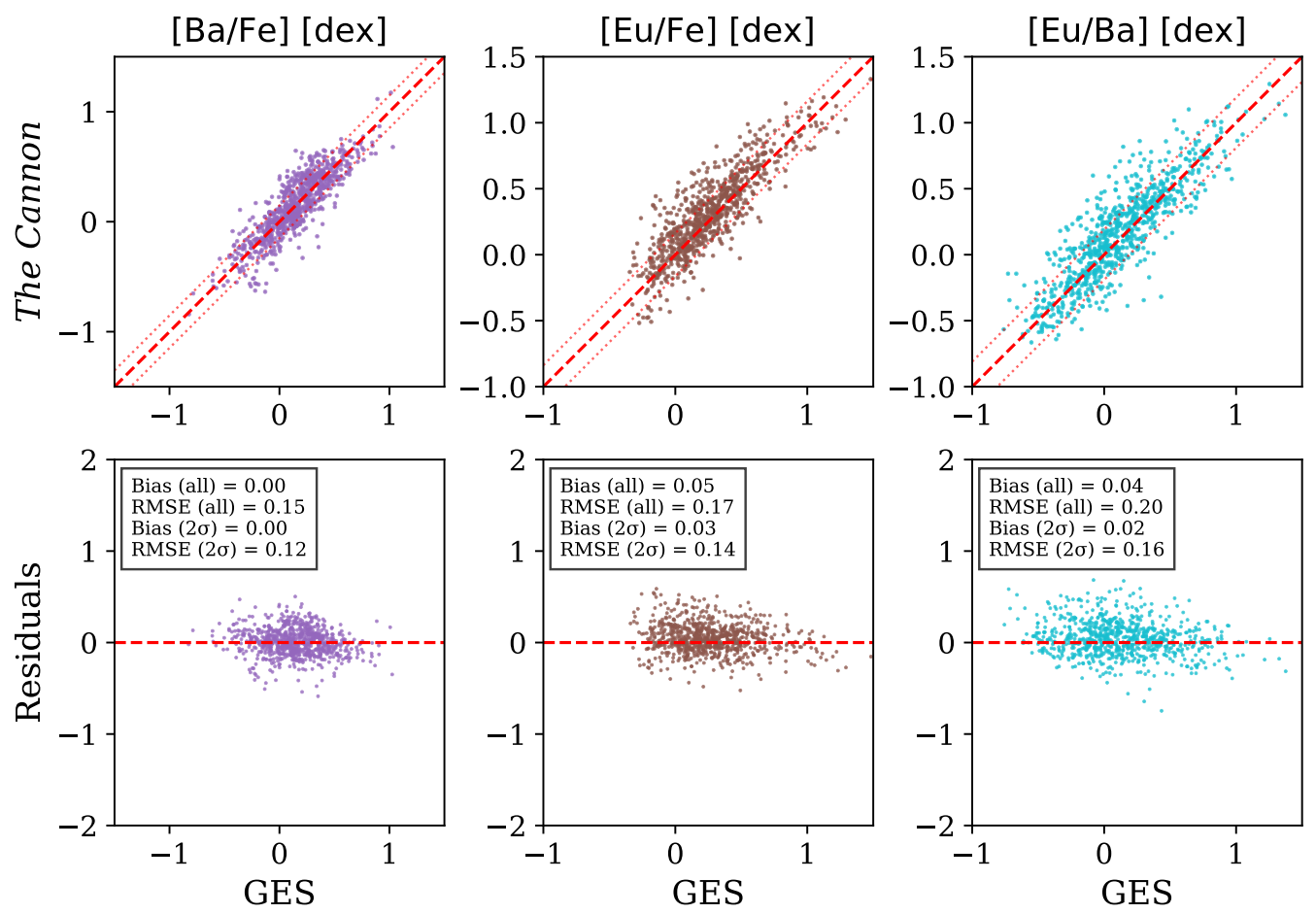

**Fig. 3.** One-to-one comparison between *The Cannon*-derived and GES reference values for the derived abundance ratios [Ba/Fe], [Eu/Fe], and [Eu/Ba] (similar to Fig. 2).

The sensitivity of the model flux at each wavelength pixel to the input stellar labels can be analysed from the distribution of model coefficients (see $\theta_\lambda$ in Sect. 3). Figure B.1 shows the linear-order model coefficients obtained using *The Cannon* for all the stellar labels. We find distinct peaks in the model-coefficients for Ba II and Eu II at 6496.9 Å and 6645.1 Å, respectively, thereby showing the relevant pixels in the spectra for these lines.

Figure 5 shows the Kiel diagram of the entire test sample of 36 263 spectra, colour-coded by *The Cannon*-derived [Fe/H]. In this sample, 23 745 stars (65.5%) satisfy `flag_atm = 0` for reliable stellar parameters, while the remaining 12 518 stars (34.5%) are classified as extrapolations with `flag_atm = 1`. These extrapolated stars correspond to regions of the parameter space that are sparsely sampled by the training set, and are primarily located toward the extremes of $T_{\rm eff}$, $\log g$, and [Fe/H]. Figures D.1a and D.1b present examples of HR15N spectra, comparing the observed spectral fluxes with the spectra generated by *The Cannon*, and showing the Ba II (6496.9 Å) and Eu II (6645.1 Å) lines, respectively. *The Cannon*-generated spectra show good agreement with the observed flux with low residual scatter, demonstrating the model's ability to learn the line features for these elements.

We also show examples of observed and *The Cannon*-generated spectra across different regions of the Kiel diagram to illustrate the model performance across the stellar parameter space. The Kiel diagrams in Figs. 6a and 6b are colour coded by [Ba/Fe] and [Eu/Fe] abundances, respectively. Rectangular boxes mark selected regions spanning the range of $T_{\rm eff}$ and $\log g$ covered by the test sample, and representative stars were chosen from these populated regions to illustrate the spectral behaviour. The inset panels show a comparison between the observed spectra and the spectra generated by *The Cannon* for the same star. The vertical dashed line marks the wavelength of the Ba II line at 6496.9 Å in Fig. 6a and the Eu II line at 6645.1 Å in Fig. 6b.

These examples demonstrate that the model successfully reproduces the observed line profiles across a wide range of stellar parameters and abundance values, indicating that the neutron-capture element abundances derived by the model are reliable across the parameter space.

### 6.1. Validation with Gaia Benchmark Stars

We cross-matched our sample with the *Gaia* Benchmark Stars (hereafter GBS; Jofré et al. 2018) and found 16 stars in common satisfying `flag_atm = 0`. We compared their stellar parameters derived with *The Cannon* ($T_{\rm eff}$, $\log g$, and [Fe/H]) with the corresponding GBS reference values. Figure 7 shows the residuals between *The Cannon*-derived labels and the GBS reference values for these 16 stars. The figure also shows the residuals relative to the corresponding GES reference values.

Overall, we find good agreement in $T_{\rm eff}$, $\log g$, and [Fe/H] for most of the GBS across the parameter space covered by the training set. For the 13 stars with [Fe/H] > −0.5 dex, *The Cannon* predictions agree well with both GBS and GES measurements. The residuals relative to GBS exhibit typical dispersions of ∼85 K in $T_{\rm eff}$, ∼0.18 dex in $\log g$, and ∼0.08 dex in [Fe/H]. Relative to GES, the residuals are similarly consistent, with dispersions of ∼95 K, ∼0.15 dex, and ∼0.18 dex in $T_{\rm eff}$, $\log g$, and [Fe/H], respectively. For the three remaining GBS stars in the lower-metallicity regime ([Fe/H] < −0.5 dex; HD102200, HD22879, and HD84937), the dispersion in [Fe/H] is higher, while residuals in $T_{\rm eff}$ and $\log g$ remain within ∼200 K and ∼0.2 dex, respectively. This is attributed to the less-representative label coverage of the training sample at lower metallicities, resulting in higher residuals for the metal-poor regime. Comparisons with GES measurements also show similar trends, confirming that these residuals are consistent across independent reference datasets.

In general, we find that *The Cannon* obtains the stellar parameters of GBS stars with high fidelity within the well-sampled regions of the training label space. Residuals are small for solar- and subsolar-metallicity stars, where the training set is densely populated. The agreement with GES values further supports the reliability of the model, as residuals display comparable patterns and dispersions.

### 6.2. Validation with GALAH DR4

The fourth data release of the GALactic Archaeology with HERMES survey (GALAH DR4; Buder et al. 2025) provides stellar parameters (e.g. $T_{\rm eff}$, $\log g$, [Fe/H]) and up to 32 elemental abundances for approximately one million stars, observed at a resolving power of $R \approx 28\,000$ with HERMES on the Anglo-Australian Telescope.

We cross-matched our sample with GALAH DR4 to compare against *The Cannon*-derived stellar labels. The primary quality cuts used for the cross-matched sample for all labels ($T_{\rm eff}$, $\log g$, [Fe/H], [Ba/Fe], and [Eu/Fe]) include `flag_sp = 0`, `flag_sp_fit = 0`, `flag_red = 0` and `snr_px_ccd3 ≥ 30` from GALAH DR4, and also satisfy `flag_atm = 0` from this work. Following this, we also applied additional flags: for Ba, we used `flag_ba_fe = 0` from GALAH, and `flag_Ba = 0` and `3-sigma_Ba = 0` from this work; and similarly for Eu, we used `flag_eu_fe = 0` from GALAH, and `flag_Eu = 0` and `3-sigma_Eu = 0` from this work.

Figure 8 shows the one-to-one correspondence between *The Cannon*-derived labels and the corresponding values from GALAH DR4. We find good agreement between the two datasets, with the scatter quantified using the median absolute deviation. The measured dispersions are ~100 K for $T_{\rm eff}$, ~0.2 dex for $\log g$, ~0.07 dex for [Fe/H], ~0.3 dex for [Ba/Fe], and ~0.2 dex for [Eu/Fe]. One of the primary contributions to this scatter can be attributed to the inherent offsets between the data from GALAH and our training reference, GES (see e.g. Table 4 in Hourihane et al. 2023). These offsets depend on the spectroscopic techniques employed, the line lists adopted for the abundance determinations, and cross-survey differences.

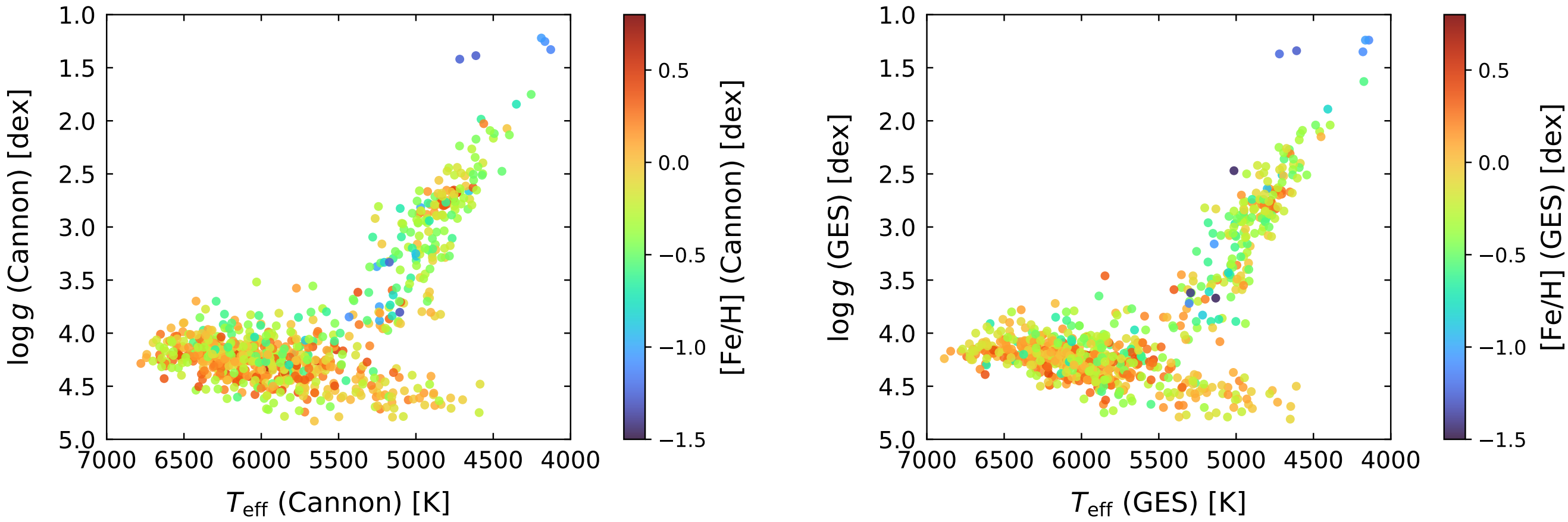

**Fig. 4.** Kiel diagrams of the training sample (782 stars). *Left*: stellar parameters predicted by *The Cannon* from the HR15N spectra, showing the effective temperature ($T_{\rm eff}$) and surface gravity ($\log g$). *Right*: reference stellar parameters from the GES catalogue for the same stars in the training sample. In both panels, the points are colour-coded by the metallicity [Fe/H] predicted by *The Cannon* and GES, respectively.

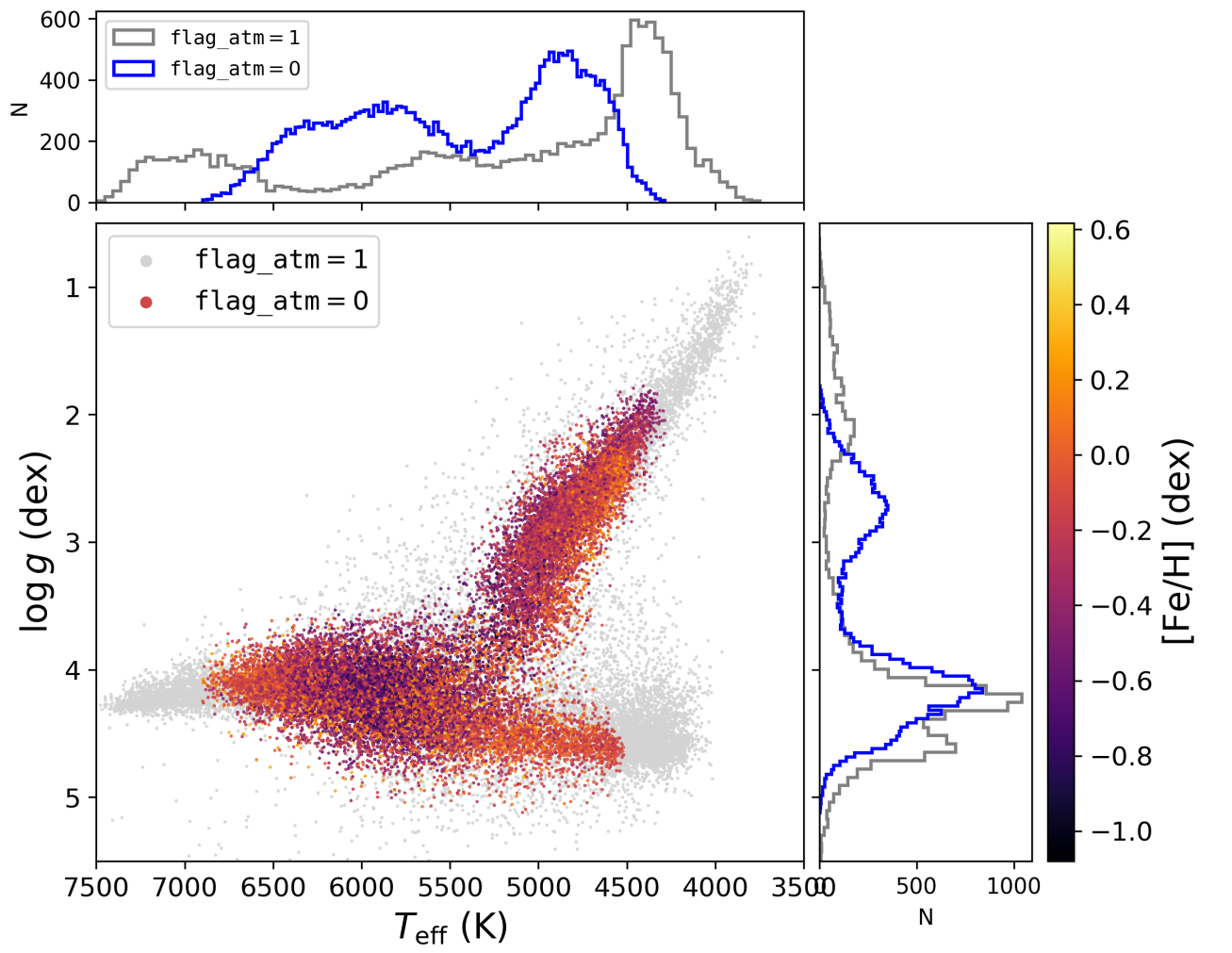

**Fig. 5.** Kiel diagram for the test sample using *The Cannon*-derived labels ($T_{\rm eff}$, $\log g$, and [Fe/H]). Stars with labels within the limits of the training sample's label space (`flag_atm = 0`) are shown as coloured dots, where the colour scale indicates values for [Fe/H]. Sources flagged by the model (`flag_atm = 1`) are plotted in light grey and refer to the stars with labels outside the label space of the training sample. The histogram distributions for both of these samples in $T_{\rm eff}$ and $\log g$ are also plotted above and to the right of the Kiel diagram, respectively.

We note that the uncertainties derived from the internal five-fold cross-validation quantify the precision of the model on the GES reference scale, demonstrating internal consistency in the predicted labels. While these estimates are anchored to a homogeneous set of training labels, comparisons with independent surveys provide a complementary validation of the results by incorporating differences in observational set-ups and analysis methodologies. The scatter observed in the comparison with GALAH DR4 is therefore consistent with the combined effects of cross-survey systematics and methodological differences, rather than being dominated by the intrinsic limitations of the model.

### 6.3. Validation with open clusters

We validated our derived abundances for several example open clusters present in our final catalogue by comparing them with values reported in the literature. Tables 2 and 3 list the mean [Ba/Fe] and [Eu/Fe] abundance ratios for these clusters, respectively, along with the number of stars used to compute the error-weighted mean abundances ($N$) and the corresponding literature references for the reported abundances. The stars used to compute the weighted mean abundances satisfy the following criteria: `flag_atm = 0`, `flag_X = 0`, and `3-sigma_X = 0` for $X$ = `Ba` or `Eu`; and cluster membership probability (`MEMQG`, Jackson et al. 2022) ≥`0.99`. Table E.1 presents *The Cannon*-derived metallicities ([Fe/H]) and neutron-capture element abundances ($A$(Ba), [Ba/Fe], $A$(Eu), [Eu/Fe]) for the open clusters in the *Gaia*-ESO HR15N sample.

A key factor in interpreting comparisons of open-cluster abundances with literature values is the number of stars contributing to the cluster mean, together with its associated uncertainty. For some clusters, particularly in the case of Eu, the number of stars contributing to the mean abundance is small ($N \approx$ 2–4). While these sample sizes limit how representative the resulting means are of the full cluster abundance distribution, the quoted uncertainties on the weighted means reflect the individual stellar uncertainties and the number of contributing stars, and the mean abundances remain broadly consistent with literature measurements.

For NGC 2516, we derive a mean [Ba/Fe] of 0.17 ± 0.01 dex, in good agreement with the value reported by Magrini et al. (2018) (0.20 ± 0.08 dex), while being lower than earlier measurements by D'Orazi et al. (2009) (0.41 ± 0.04 dex), Baratella et al. (2021) (0.31 ± 0.03 dex), and Spina et al. (2021) (0.30 ± 0.04 dex). In NGC 2420, our mean [Ba/Fe] of 0.12 ± 0.01 dex is slightly lower than the value reported in Magrini et al. (2018) (0.23 ± 0.09 dex). For NGC 6705, we obtain 0.20 ± 0.03 dex, in agreement within the quoted uncertainties of 0.15 ± 0.12 dex (Magrini et al. 2018). For Ruprecht 134, we measure −0.26 ± 0.07 dex, which overlaps with the spread of

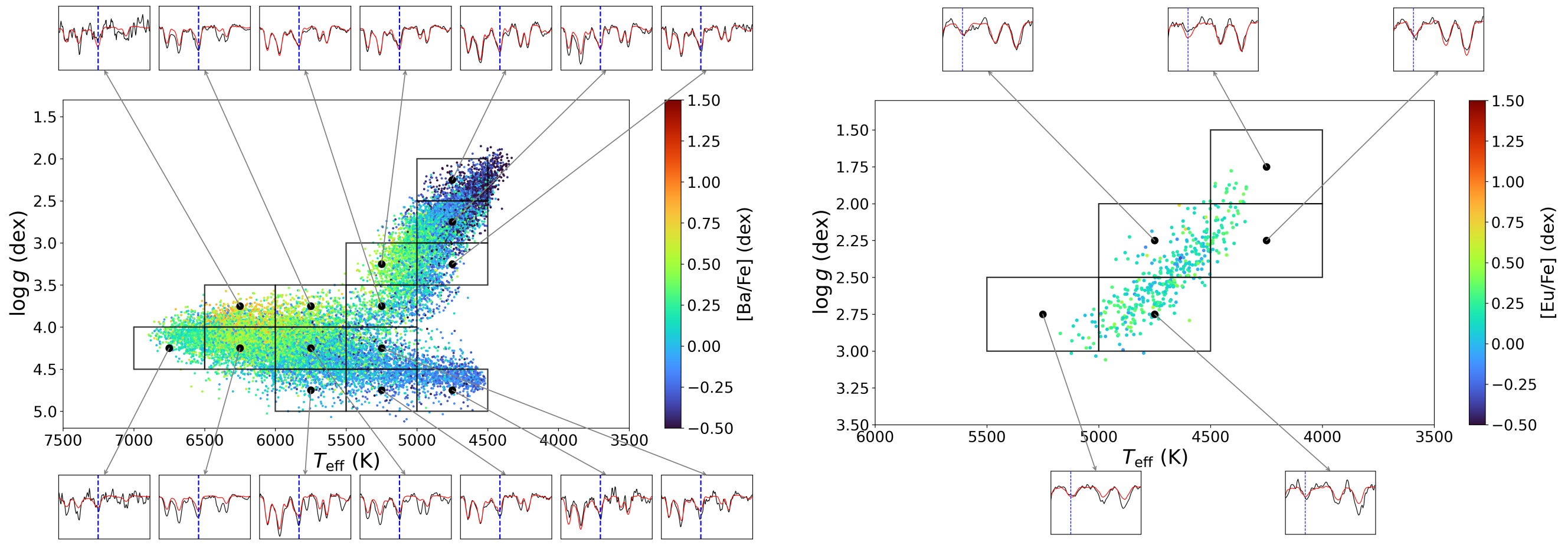

(a) [Ba/Fe] distribution across the stellar parameter space for stars satisfying `flag_atm = 0`, `flag_Ba = 0`, and `3-sigma_Ba = 0`.

(b) [Eu/Fe] distribution across the stellar parameter space for stars satisfying `flag_atm = 0`, `flag_Eu = 0`, and `3-sigma_Eu = 0`.

**Fig. 6.** Kiel diagrams of the sample colour-coded by [Ba/Fe] and [Eu/Fe], respectively, with representative stars shown as inset spectra. Black rectangles indicate regions (boxes) defined with increments of 500 K in $T_{\rm eff}$ and 0.5 dex in $\log g$. Arrows connect each inset to its corresponding box. The inset panels show the observed HR15N spectra (black) and the corresponding *The Cannon*-generated spectra (red) for the representative stars along with the scatter, zoomed around the Ba feature at 6496.9 Å (*left* panel) and the Eu feature at 6645.1 Å (*right* panel). The model scatter (uncertainty in the generated flux) inside each inset is sufficiently small to remain visually indistinguishable in most panels.

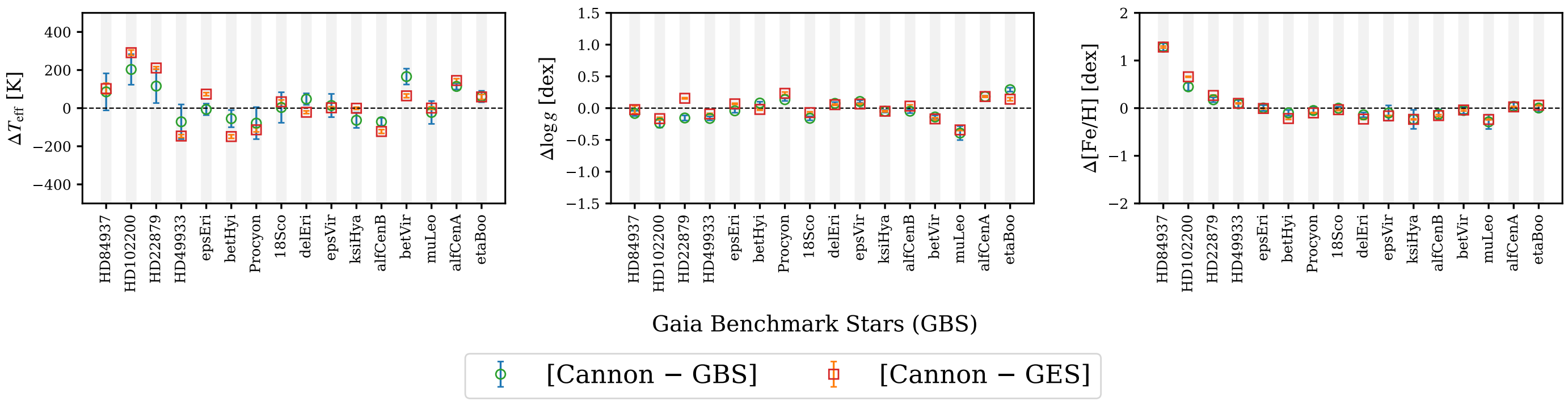

**Fig. 7.** Residuals of *The Cannon*-derived labels ($T_{\rm eff}$, $\log g$, and [Fe/H]) relative to the GBS and GES values. The green circles correspond to (*Cannon − GBS*) residuals, and the red squares correspond to (*Cannon − GES*) residuals. The error bars are very small, and are therefore not directly visible in the figure. The stars are ordered along the *x*-axis by increasing metallicity ([Fe/H]).

−0.15 ± 0.08 dex reported by Magrini et al. (2018). NGC 4815 shows similar mean values of 0.33 ± 0.07 dex (this work) and 0.31 ± 0.13 dex (Magrini et al. 2018) for [Ba/Fe]. For more distant and older clusters, including Berkeley 32 and Berkeley 39, our results remain consistent with previous determinations. In Berkeley 32, we find 0.25 ± 0.02 dex, within the collective range 0.22–0.51 dex reported in the literature references (Bragaglia et al. 2008; D'Orazi et al. 2009; Carrera & Pancino 2011; Yong et al. 2012; Jacobson & Friel 2013; Spina et al. 2021), whereas in Berkeley 39, we obtain 0.11 ± 0.01 dex, close to the reported value of 0.07 ± 0.06 dex (Jacobson & Friel 2013).

For [Eu/Fe], our derived abundances are in overall agreement with both GES results and independent literature references. In Berkeley 32, we measure 0.21 ± 0.01 dex, comparable to 0.20 ± 0.08 dex (Van der Swaelmen et al. 2023) and 0.31 ± 0.11 dex (Yong et al. 2012) within reported spread, and higher than 0.03 ± 0.08 dex (Jacobson & Friel 2013) and 0.08 ± 0.07 dex (Spina et al. 2021). For NGC 2425, our mean of 0.18 ± 0.04 dex is consistent with 0.15 ± 0.07 dex reported by Van der Swaelmen et al. (2023). In Berkeley 21, we obtain 0.15 ± 0.02 dex, in agreement with 0.16 ± 0.05 dex (Van der Swaelmen et al. 2023) and lower than 0.31 ± 0.08 dex (Yong et al. 2012). Czernik 24 exhibits 0.12 ± 0.01 dex, in good agreement with 0.17 ± 0.03 dex reported by Van der Swaelmen et al. (2023). For Berkeley 36, we obtained a mean [Eu/Fe] of 0.21 ± 0.01 dex, consistent with 0.21 ± 0.07 dex (Van der Swaelmen et al. 2023) and within the spread of 0.13 ± 0.20 dex in Magrini et al. (2018). Finally, NGC 2420 shows similar [Eu/Fe] values of 0.12 ± 0.04 dex (this work) and 0.16 ± 0.08 dex (Van der Swaelmen et al. 2023).

We note that the literature studies mentioned for Ba and Eu abundances in open clusters (Tables 2 and 3) employ a variety of observational set-ups and analysis techniques, which may introduce systematic differences between measurements. Solar reference abundances also vary slightly between studies, with typical values of $A$(Ba) ≈ 2.17–2.18 dex (Grevesse et al. 2007; Asplund et al. 2009; Lodders et al. 2009) and $A$(Eu) ≈ 0.52 dex (Grevesse et al. 2007). D'Orazi et al. (2009) analysed Ba abundances in NGC 2516 using high-resolution UVES ($R$ ≈ 47 000) spectroscopy, deriving abundances from two Ba II lines (5853.7 and 6496.9 Å). The open-cluster abundances reported by Magrini et al. (2018) are based on the high-resolution ($R$ ≈ 47 000) UVES

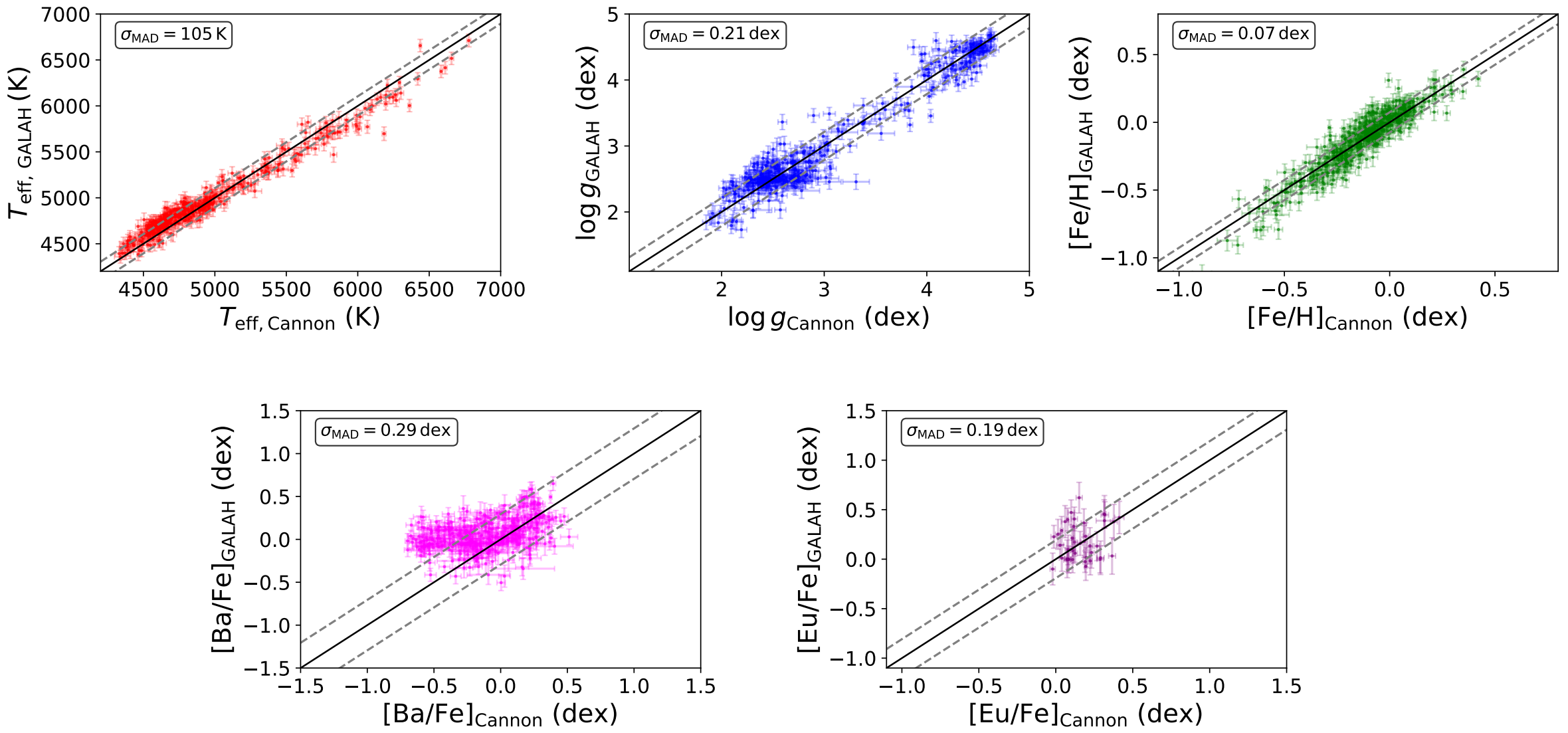

**Fig. 8.** Comparison of stellar parameters and neutron-capture element abundances between *The Cannon*-derived values and GALAH DR4 measurements. The comparisons include $T_{\rm eff}$, $\log g$, and metallicity [Fe/H], together with the abundance ratios [Ba/Fe] and [Eu/Fe]. The solid black line indicates the one-to-one relation, while the dashed grey lines show $\pm\sigma_{\rm MAD}$, where $\sigma_{\rm MAD}$ is the scaled median absolute deviation of the residuals.

**Table 2.** Mean [Ba/Fe] abundance ratios obtained for a few example open clusters.

| Cluster | [Ba/Fe] (dex) | Source |
|---|---|---|
| NGC 2516 | 0.17 ± 0.01 | This work ($N$ = 125) |
| | 0.41 ± 0.04 | D'Orazi et al. (2009) |
| | 0.20 ± 0.08 | Magrini et al. (2018) |
| | 0.31 ± 0.03 | Baratella et al. (2021) |
| | 0.30 ± 0.04 | Spina et al. (2021) |
| NGC 2420 | 0.12 ± 0.01 | This work ($N$ = 311) |
| | 0.23 ± 0.09 | Magrini et al. (2018) |
| NGC 6705 | 0.20 ± 0.03 | This work ($N$ = 49) |
| | 0.15 ± 0.12 | Magrini et al. (2018) |
| Ruprecht 134 | −0.26 ± 0.07 | This work ($N$ = 27) |
| | −0.15 ± 0.08 | Magrini et al. (2018) |
| NGC 4815 | 0.33 ± 0.07 | This work ($N$ = 16) |
| | 0.31 ± 0.13 | Magrini et al. (2018) |
| Berkeley 32 | 0.25 ± 0.02 | This work ($N$ = 160) |
| | 0.29 ± 0.10 | Bragaglia et al. (2008) |
| | 0.24 ± 0.15 | D'Orazi et al. (2009) |
| | 0.51 ± 0.12 | Carrera & Pancino (2011) |
| | 0.29 ± 0.12 | Yong et al. (2012) |
| | 0.22 ± 0.04 | Jacobson & Friel (2013) |
| | 0.45 ± 0.06 | Spina et al. (2021) |
| Berkeley 39 | 0.11 ± 0.01 | This work ($N$ = 393) |
| | 0.07 ± 0.06 | Jacobson & Friel (2013) |

**Notes.** $N$ represents the number of member stars used to calculate the mean abundances weighted by their uncertainties.

**Table 3.** Mean [Eu/Fe] abundance ratios obtained for a few example open clusters (similar to Table 2).

| Cluster | [Eu/Fe] (dex) | Source |
|---|---|---|
| Berkeley 32 | 0.21 ± 0.01 | This work ($N$ = 4) |
| | 0.31 ± 0.11 | Yong et al. (2012) |
| | 0.03 ± 0.08 | Jacobson & Friel (2013) |
| | 0.08 ± 0.07 | Spina et al. (2021) |
| | 0.20 ± 0.08 | Van der Swaelmen et al. (2023) |
| NGC 2425 | 0.18 ± 0.04 | This work ($N$ = 3) |
| | 0.15 ± 0.07 | Van der Swaelmen et al. (2023) |
| Berkeley 21 | 0.15 ± 0.02 | This work ($N$ = 4) |
| | 0.31 ± 0.08 | Yong et al. (2012) |
| | 0.16 ± 0.05 | Van der Swaelmen et al. (2023) |
| Czernik 24 | 0.12 ± 0.01 | This work ($N$ = 2) |
| | 0.17 ± 0.03 | Van der Swaelmen et al. (2023) |
| Berkeley 36 | 0.21 ± 0.01 | This work ($N$ = 2) |
| | 0.13 ± 0.20 | Magrini et al. (2018) |
| | 0.21 ± 0.07 | Van der Swaelmen et al. (2023) |
| NGC 2420 | 0.12 ± 0.04 | This work ($N$ = 2) |
| | 0.16 ± 0.08 | Van der Swaelmen et al. (2023) |

**Notes.** $N$ represents the number of member stars used to calculate the mean abundances weighted by their uncertainties.

data from the *Gaia*-ESO Survey. Baratella et al. (2021) analysed high-resolution, high-S/N spectra from the FLAMES/UVES-580 set-up (4800–6800 Å) in the *Gaia*-ESO Survey for solar-type dwarf stars present in young open clusters. Here, they derived Ba abundances using the 5853.7 Å line, and further used the hyperfine structure data (McWilliam 1998) and adopted the isotopic solar mixtures for Ba. Yong et al. (2012) derived elemental abundances for stars in outer-disc open clusters using spectra ($R \approx 48\,000$) obtained with the High Resolution Echelle Spectrometer (HIRES; Vogt et al. 1994) spectrograph at the Keck I telescope, based on equivalent-width measurements and spectral synthesis of selected absorption lines. Jacobson & Friel (2013) derived Zr, Ba, La, and Eu abundances for 50 stars in 19 open clusters using spectra from Hydra/WIYN and HIRES/Keck ($R \approx 18\,000$–$28\,000$, and wavelength coverage ∼5000–8100 Å), measuring lines for these elements with equivalent widths and spectral synthesis, including hyperfine and isotopic structure where appropriate. Spina et al. (2021) used chemical abundances from the GALAH Survey, based on high-resolution ($R \approx 28\,000$) HERMES spectra (Brzeski et al. 2011; Heijmans et al. 2012) obtained at the Anglo-Australian Telescope over four wavelength bands (~4713–4903, 5648–5873, 6478–6737, and 7585–7887 Å), with elemental abundances derived through the GALAH analysis pipeline. Carrera & Pancino (2011) obtained high-resolution ($R \approx 30\,000$), high-S/N spectra from the fibre spectrograph FOCES (Pfeiffer et al. 1998) on the 2.2 m telescope at Calar Alto Observatory, for the stars in four open clusters (Hyades, Praesepe, Berkeley 32, NGC 752), deriving Ba abundances via equivalent-width analysis among 16 elements. Van der Swaelmen et al. (2023) used Eu abundances from the *Gaia*-ESO iDR6 Survey, based on high-resolution FLAMES/UVES spectra ($R \approx 47\,000$) obtained with the UVES-580 set-up covering the wavelength range from 4800–6800 Å.

These methodological differences in spectral resolution, wavelength coverage, line selection, and analysis techniques may lead to systematic offsets when comparing measurements across studies. Collectively, these comparisons demonstrate that our derived [Ba/Fe] and [Eu/Fe] abundances are broadly consistent with previously published measurements, supporting the analysis across both s- and r-process neutron capture abundances in open clusters.

## 7. Conclusions

We have demonstrated the application of *The Cannon* data-driven spectral inference framework to the *Gaia*-ESO Survey spectroscopic dataset, deriving a catalogue of homogeneous stellar parameters ($T_{\rm eff}$, $\log g$, $v \sin i$, [Fe/H]) and neutron-capture elemental abundances ($A$(Ba) and $A$(Eu)) for 36 263 spectra observed with the FLAMES/GIRAFFE HR15N set-up. The model was trained on a selected sample of 782 high-fidelity HR15N spectra using the reference labels from the *Gaia*-ESO homogenised catalogue (Hourihane et al. 2023), ensuring that all the inferred labels remain on the same scale as the *Gaia*-ESO reference labels.

Calibrated distance metrics in multi-dimensional label space – $D_{\rm atm}$ for the four stellar parameters ($T_{\rm eff}$, $\log g$, $v \sin i$, and [Fe/H]), and $D_{\rm Ba}$ and $D_{\rm Eu}$ for the element-specific label spaces – provide quantitative separation between interpolation and extrapolation regimes, as encoded by `flag_atm` for the stellar parameters and `flag_Ba`/`flag_Eu` for the element-specific abundances $A$(Ba) and $A$(Eu). These thresholds are calibrated from the training-sample's local-distance distributions, with each label dimension normalised by its five-fold cross-validation RMSE, and provide a robust framework for identifying reliable stellar label measurements. Complementary per-pixel line-detection criteria (`3-sigma_X = 0`, for $X$ = Ba or Eu) require the corresponding spectral features to satisfy the adopted detection thresholds for each abundance determination. Together, these criteria provide internal diagnostics – including label distances, element-specific reduced $\chi^2$ values, and $3\sigma$ detection flags – for *The Cannon* predictions, thereby yielding a catalogue suitable for quantitative Galactic population studies.

We computed the total uncertainties for each label by combining the internal model uncertainties ($\sigma_{\rm int}$), extracted from the label-space covariance matrix, with the S/N-dependent dispersions ($\sigma_{\rm obs}$) derived from 100 noise realisations per spectrum in quadrature, as detailed in Sect. 5. The median values of total uncertainties in the sample satisfying `flag_atm = 0` are ~38 K for $T_{\rm eff}$, ~0.08 dex for $\log g$, and ~0.04 dex for [Fe/H]. Similarly, for stars satisfying `flag_X = 0` and `3-sigma_X = 0` for X = Ba and Eu, the median uncertainties are ~0.06 dex and ~0.02 dex for $A$(Ba) and $A$(Eu), respectively. We also provide the relative abundances [Ba/Fe] and [Eu/Fe] in the final catalogue, along with their associated total uncertainties.

External validation against *Gaia* Benchmark Stars (Jofré et al. 2018) shows generally small residuals across the label space spanned by the training set, with larger residuals for stars located in less densely sampled regions of the training-label space. Cross-survey comparison with GALAH DR4 (Buder et al. 2025) shows good agreement across all labels, with typical dispersions of ~100 K in $T_{\rm eff}$, ~0.2 dex in $\log g$, ~0.07 dex in [Fe/H], ~0.3 dex in [Ba/Fe], and ~0.2 dex in [Eu/Fe]. These dispersions reflect the combined effects of differing instrumental set-ups, cross-survey offsets, adopted spectroscopic pipelines, and corresponding calibration strategies. Validation using several open clusters shows that the mean [Ba/Fe] and [Eu/Fe] ratios are consistent with literature values, supporting a reliable abundance scale for both s- and r-process neutron-capture elements.

Our final catalogue provides 21 055 reliable Ba abundances (`flag_Ba = 0` and `3-sigma_Ba = 0`) and 409 reliable Eu abundances (`flag_Eu = 0` and `3-sigma_Eu = 0`). The smaller number of reliable Eu abundances compared to Ba primarily reflects the intrinsically weaker Eu II 6645.1 Å feature at the HR15N resolving power, which requires higher S/N and favourable stellar parameters for reliable detection. Consequently, fewer stars satisfy the adopted $\geq 3\sigma$ detection criterion for Eu relative to the stronger Ba II 6496.9 Å line. Among the derived abundances, 10 353 Ba and 133 Eu measurements are newly reported in this work, extending the available neutron-capture abundance information beyond the existing *Gaia*-ESO and GALAH DR4 catalogues and increasing the number of stars with available Ba and Eu abundance measurements from HR15N spectra. The catalogue therefore constitutes a homogeneous dataset of Ba and Eu abundances spanning a broad range of open-cluster ages and Galactic environments, enabling quantitative studies of age- and galactocentric radius-dependent neutron-capture enrichment. Further analysis of the neutron-capture elemental abundance profiles for the open clusters in the sample is discussed in Das et al. (in prep).

Future work could incorporate NLTE corrections for the Ba II and Eu II lines analysed here. For the Ba II 6496.9 Å transition, NLTE grids such as those of Korotin et al. (2015) demonstrate that the abundance corrections depend on effective temperature, surface gravity, and metallicity, and can reach values of ~0.1 dex depending on stellar parameters. For the Eu II 6645.1 Å line, NLTE modelling (Guo et al. 2025) indicates generally small but non-zero corrections, with a dependence on atmospheric structure and metallicity. Therefore, incorporating these NLTE calculations would facilitate a systematic exploration of their effects on the [Eu/Ba] ratios.

Owing to the uniform analysis framework adopted at HR15N spectral resolution, the methodology developed in this work can be readily extended to spectroscopic data from forthcoming wide-field facilities such as 4MOST (de Jong et al. 2019), WEAVE (Dalton et al. 2018), and WST (Bacon et al. 2024). This provides a scalable framework for efficiently deriving neutron-capture abundances in large stellar samples, enabling future studies of Galactic chemical evolution across significantly larger and more diverse stellar populations.

## Data availability

The full *Cannon*-GES catalogue (Table A.1) is available at the CDS via https://cdsarc.cds.unistra.fr/viz-bin/cat/J/A+A/713/A148.

*Acknowledgements.* P.B.D. acknowledges support from the Early-Career Scientific Visitor Programme at the European Southern Observatory (ESO), Garching, under which this research was carried out. P.B.D. also acknowledges support from the Australian Commonwealth Government Research Training Program (*RTP*) Scholarship (https://doi.org/10.82133/C42F-K220). G.G. acknowledges support by Deutsche Forschungsgemeinschaft (DFG, German Research Foundation) – project-IDs: eBer-22-59652 (GU 2240/1-1 "Galactic Archaeology with Convolutional Neural-Networks: Realising the potential of *Gaia* and 4MOST"). This project has received funding from the European Research Council (*ERC*) under the European Union's Horizon 2020 research and innovation programme (Grant agreement No. 949173). The authors thank C.C. Worley and M.J. Irwin for suggestions with the data analysed in this work. P.B.D. and F.P. also thank Martino Romaniello for the useful discussions and insights into spectral processing and various systematics of the *Gaia*-ESO data. This work is based on observations collected with ESO telescopes at the La Silla Paranal Observatory in Chile, for the *Gaia*-ESO Large Public Spectroscopic Survey (188.B-3002, 193.B-0936, 197.B-1074). This research has made use of data obtained from the *GES* Data Archive, prepared and hosted by the Wide Field Astronomy Unit, Institute for Astronomy, University of Edinburgh, which is funded by the UK Science and Technology Facilities Council. This work has made use of data from the European Space Agency (ESA) mission *Gaia* (https://www.cosmos.esa.int/gaia), processed by the *Gaia* Data Processing and Analysis Consortium (DPAC, https://www.cosmos.esa.int/web/gaia/dpac/consortium). Funding for the DPAC has been provided by national institutions, in particular the institutions participating in the *Gaia* Multilateral Agreement. This research also used the NASA Astrophysics Data System (*ADS*) bibliographic services and the following software packages: `Astropy` (http://www.astropy.org) (Astropy Collaboration 2013, 2018), `NumPy` (Harris et al. 2020), `pandas` (McKinney 2010), `Matplotlib` (Hunter 2007), and `TOPCAT` (Taylor 2005, 2011).

## Appendix A: Table schema of the final catalogue

The final catalogue presented in this work provides *The Cannon*-derived stellar parameters and elemental abundances for 36 263 HR15N spectra. It includes the source identifier (`ObjID`) and catalogue metadata (`GES_FLD`, RA, and DEC) adopted from the GES DR5 catalogue (Hourihane et al. 2023). The stellar parameters ($T_{\rm eff}$, $\log g$, $v \sin i$, and [Fe/H]) are accompanied by reliability flags and label distances. The abundances $A$(Ba) and $A$(Eu), together with the abundance ratios [Ba/Fe] and [Eu/Fe], are reported with their respective uncertainties, associated quality flags, and reduced $\chi^2$ fit metrics for the corresponding spectral windows.

Table A.1: Table schema for *The Cannon*-GES catalogue in this work.

| Column | Unit | Description |
|---|---|---|
| ObjID | – | *Gaia*-ESO object identifier |
| GES_FLD | – | *Gaia*-ESO field name |
| RA | deg | Right Ascension |
| DEC | deg | Declination |
| S/N | – | Signal-to-noise ratio of spectra |
| rchi_sq | – | Reduced $\chi^2$ value for *The Cannon*-fitted spectra |
| Teff | K | *The Cannon*-derived effective temperature |
| e_Teff | K | Uncertainty in effective temperature |
| logg | dex | *The Cannon*-derived surface gravity |
| e_logg | dex | Uncertainty in surface gravity |
| vsini | km s$^{-1}$ | *The Cannon*-derived projected rotational velocity |
| e_vsini | km s$^{-1}$ | Uncertainty in projected rotational velocity |
| FeH | dex | *The Cannon*-derived [Fe/H] |
| e_FeH | dex | Uncertainty in [Fe/H] |
| flag_atm | – | Distance-based quality flag for stellar parameters (0/1) |
| atm_dist | – | Label distance ($D_{\rm atm}$) for stellar parameters |
| A(Ba) | dex | *The Cannon*-derived A(Ba) |
| e_A(Ba) | dex | Uncertainty in A(Ba) |
| BaFe | dex | *The Cannon*-derived [Ba/Fe] |
| e_BaFe | dex | Uncertainty in [Ba/Fe] |
| flag_Ba | – | Distance-based quality flag for Ba (0/1) |
| Ba_dist | – | Label distance ($D_{\rm Ba}$) for A(Ba) |
| 3-sigma_Ba | – | Flag for Ba detection at 3$\sigma$ significance (0/1) |
| rchi_sq_Ba | – | Reduced $\chi^2$ value for *The Cannon*-fitted spectra in Ba wavelength window |
| A(Eu) | dex | *The Cannon*-derived A(Eu) |
| e_A(Eu) | dex | Uncertainty in A(Eu) |
| EuFe | dex | *The Cannon*-derived [Eu/Fe] |
| e_EuFe | dex | Uncertainty in [Eu/Fe] |
| flag_Eu | – | Distance-based quality flag for Eu (0/1) |
| Eu_dist | – | Label distance ($D_{\rm Eu}$) for A(Eu) |
| 3-sigma_Eu | – | Flag for Eu detection at 3$\sigma$ significance (0/1) |
| rchi_sq_Eu | – | Reduced $\chi^2$ value for *The Cannon*-fitted spectra in Eu wavelength window |

**Notes.** A flag value of "0" indicates that the corresponding quality or detection criterion is satisfied (i.e. the flag is not problematic).

We recommend that users adopt quality flags according to the desired scientific application. For studies involving stellar parameters, we recommend selecting sources with `flag_atm = 0`. For analyses of elemental abundances $A(X)$, where $X$ denotes Ba or Eu, we recommend using `flag_X = 0` and `3-sigma_X = 0`. The `flag_X` criterion indicates that the abundance lies within the label space sampled by the training set, while the `3-sigma_X` criterion corresponds to detection of the spectral feature at the adopted significance threshold. The columns `atm_dist`, `Ba_dist`, and `Eu_dist` are also provided to allow users to impose more conservative or more inclusive cuts depending on the requirements of their analysis.

For analyses of relative abundance ratios [$X$/Fe], we recommend requiring the combined criteria `flag_atm = 0`, `flag_X = 0`, and `3-sigma_X = 0`. The abundance ratios `BaFe` and `EuFe` are calculated from $A$(Ba), $A$(Eu), and [Fe/H] derived with *The Cannon*, using the solar reference values adopted by the *Gaia*-ESO Survey (see Eqs. 11 and 12). Their uncertainties, `e_BaFe` and `e_EuFe`, include the propagated uncertainties from both the elemental abundance and metallicity. The reduced $\chi^2$ values provide additional diagnostics of both the global spectral fit and the local fit quality within the spectral windows centred on the Ba and Eu absorption features at 6496.9 Å and 6645.1 Å, respectively.

## Appendix B: *The Cannon* model linear coefficients for all stellar labels

The model coefficients (Fig. B.1) provide insight into how *The Cannon* maps variations in stellar labels to specific spectral features. They indicate the wavelengths sensitive to each parameter, with $A$(Ba) and $A$(Eu) showing responses at the expected Ba II and Eu II lines, demonstrating that the model successfully identifies key diagnostic regions. The trends for $T_{\rm eff}$, $\log g$, and [Fe/H] reflect the broader sensitivity of the entire spectrum to fundamental stellar properties.

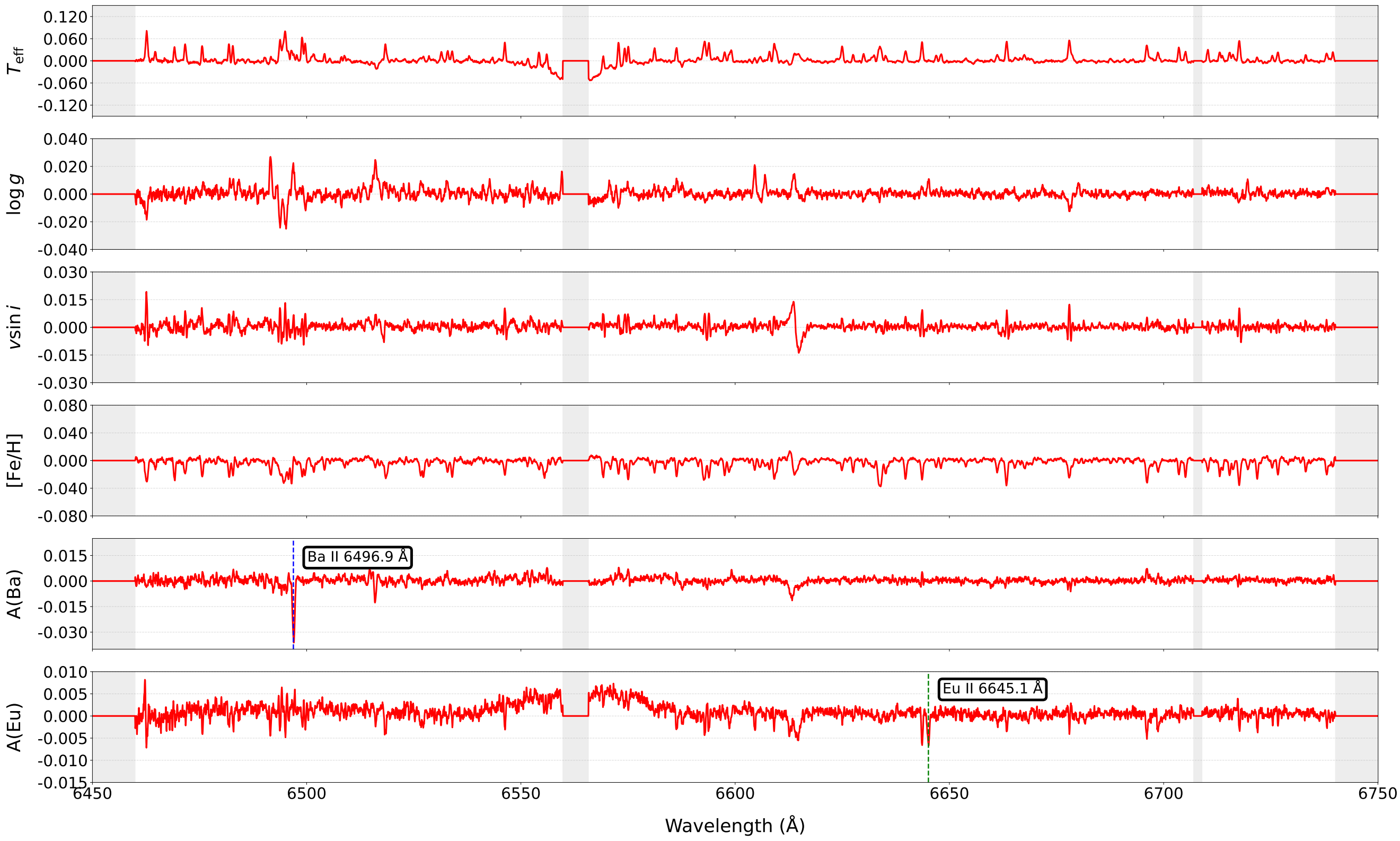

Fig. B.1: Linear coefficients optimised by *The Cannon* model for the trained sample over the HR15N wavelength range. Each panel shows the variations of the first-order coefficients for the stellar labels: $T_{\rm eff}$, $\log g$, $v \sin i$, [Fe/H], $A$(Ba), and $A$(Eu) across the entire spectrum. The grey shaded regions indicate masked spectral intervals, including ±3 Å around H$\alpha$ at 6562.8 Å, the blue and red wavelength edges (10 Å at both ends), and the Li I line at 6707.8 Å in the wavelength window (6707–6709 Å). *The Cannon* successfully identifies the pixels most sensitive to the Ba II and Eu II lines in the wavelength range, consistent with 6496.9 Å and 6645.1 Å, respectively. These lines are highlighted with vertical dashed lines and are labelled in boxes.

## Appendix C: Quality flags for Ba and Eu

Table C.1: Summary of parameter availability and quality flags in the final catalogue of 36 263 spectra.

| Label (X) | `flag_X = 0` | `3-sigma_X = 0` | `flag_X = 0` & `3-sigma_X = 0` | `flag_X = 1` & `3-sigma_X = 0` |
|---|---|---|---|---|
| Ba | 22 374 | 30 998 | 21 055 | 9 943 |
| Eu | 22 807 | 831 | 409 | 422 |

Table C.1 summarises the number of stars satisfying the quality criteria adopted in the final catalogue for Ba and Eu. The recommended Ba sample contains 21,055 stars satisfying both `flag_Ba = 0` and `3-sigma_Ba = 0`, providing a large sample for statistical investigations across the HR15N dataset. In contrast, the Eu sample contains 409 stars meeting the corresponding criteria, primarily reflecting the intrinsically weaker Eu II 6645.1 Å feature and the consequently smaller number of high-significance Eu detections. The difference between the number of stars satisfying `3-sigma_X = 0` and those in the final recommended sample highlights the complementary role of the adopted quality criteria. While `3-sigma_X` identifies statistically significant Ba and Eu line detections, `flag_X` additionally ensures that the inferred abundances are derived within regions of stellar-label space reliably represented by the training sample. The catalogue therefore provides both quality flags and associated label-distance metrics, enabling customised sample selection for different scientific applications.

## Appendix D: Comparison of HR15N observed spectra against *The Cannon*-generated model spectra

Figure D.1 presents representative examples of HR15N spectra from the analysed sample together with the corresponding *The Cannon*-generated model spectra in the wavelength regions surrounding the Ba II and Eu II features.

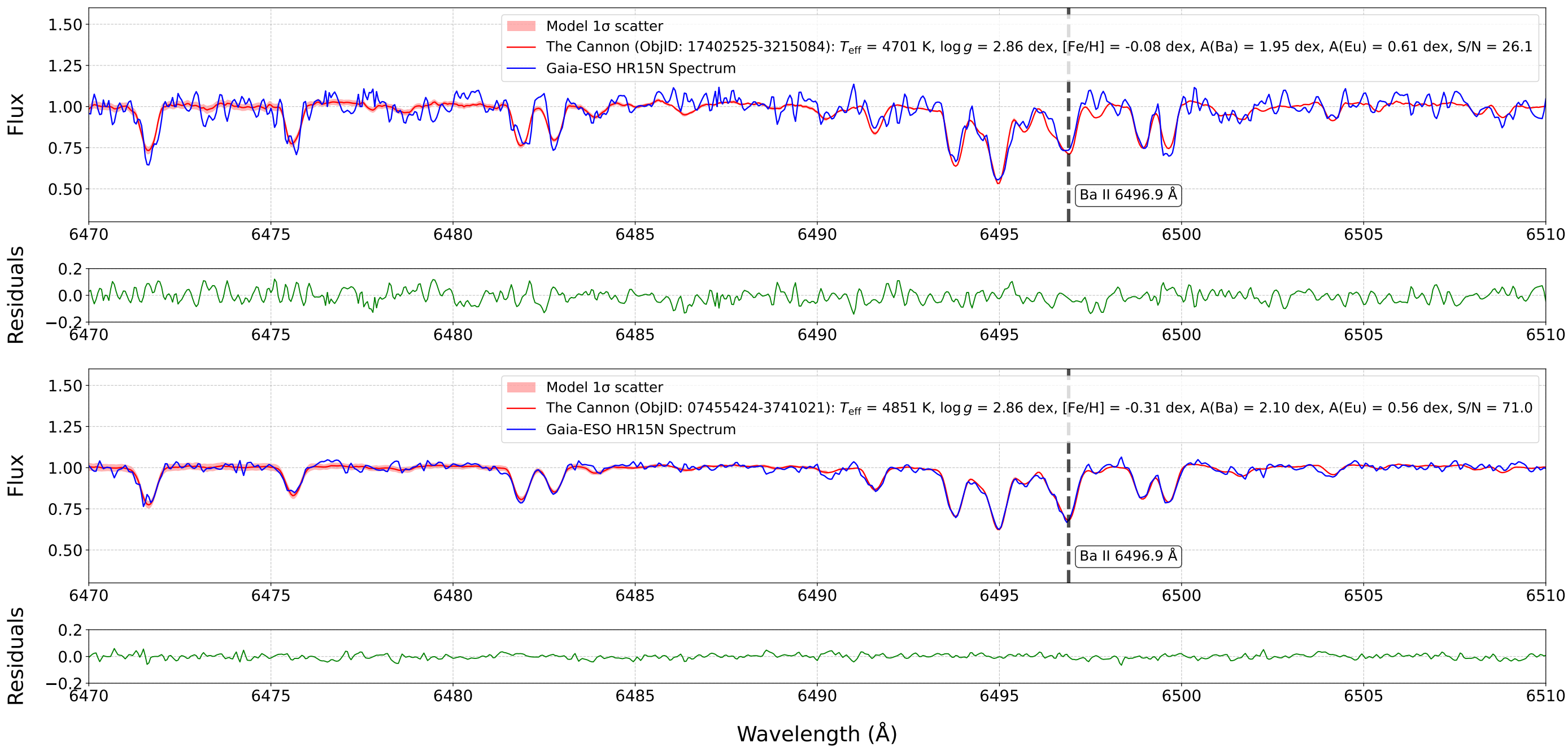

(a)

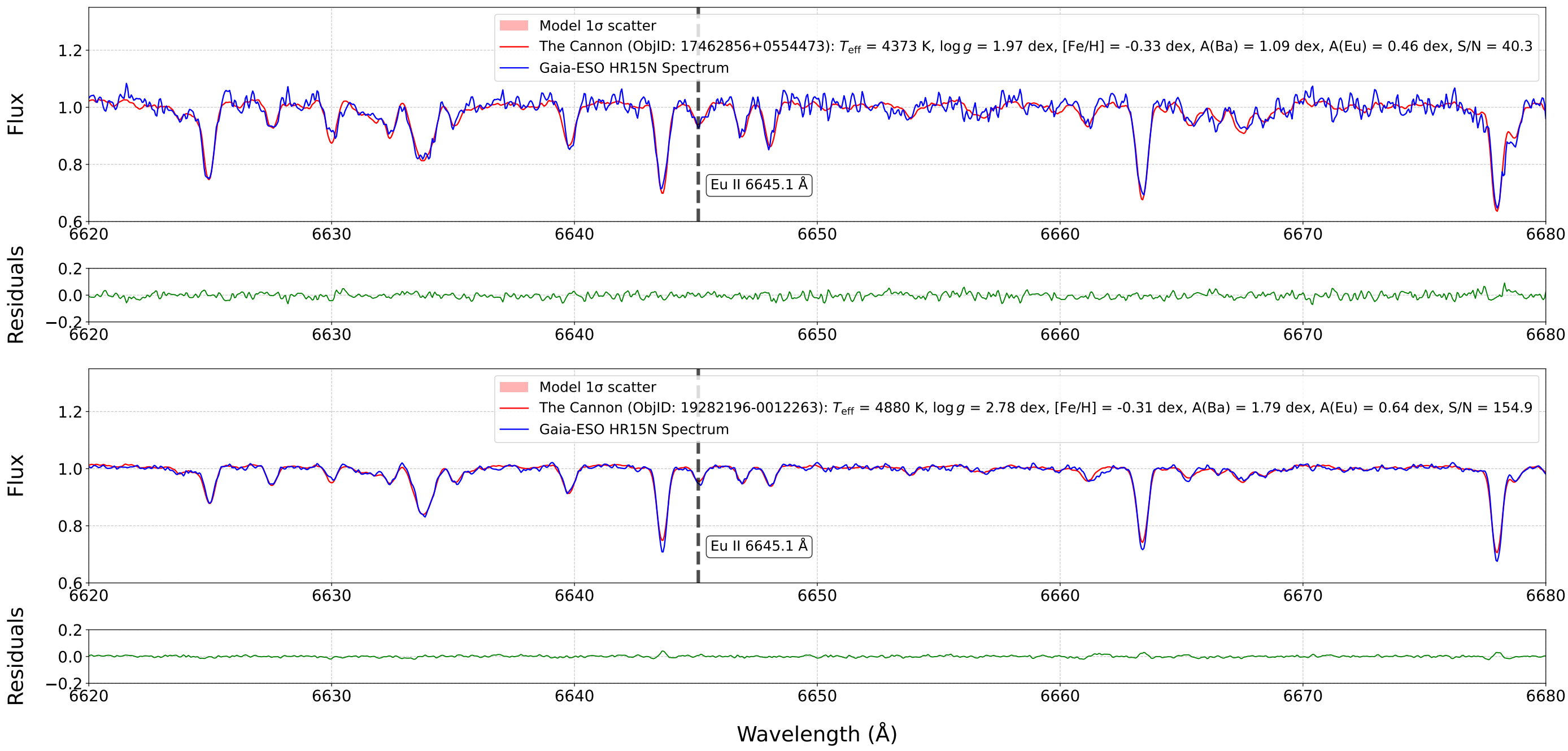

(b)

Fig. D.1: Comparison of observed HR15N spectra (blue) from the *Gaia*-ESO Survey with *The Cannon*-generated spectra (red) for example stars from our sample. The `ObjID`s and the corresponding *The Cannon*-derived stellar labels are provided in the legends of each spectrum. Each star is shown in two stacked panels: the top panel displays the spectrum itself, while the bottom panel shows the residuals between *The Cannon*-predicted and observed flux. Panel (a) highlights the Ba II absorption line at 6496.9 Å, and panel (b) highlights the Eu II line at 6645.1 Å. Residuals remain small, demonstrating that *The Cannon* reproduces the spectral features for Ba II and Eu II in these examples.

## Appendix E: *The Cannon*-derived abundances for open clusters in the *Gaia*-ESO Survey

Table E.1 presents the cluster-level summary of *The Cannon*-derived abundances for open clusters in the *Gaia*-ESO HR15N sample. For each cluster, we report mean values for [Fe/H], $A$(Ba), [Ba/Fe], $A$(Eu), and [Eu/Fe], along with their uncertainties.

Table E.1: *The Cannon*-derived mean chemical abundances for open clusters.

| Cluster | [Fe/H] (dex) | $A$(Ba)[a] (dex) | [Ba/Fe][b] (dex) | $A$(Eu)[a] (dex) | [Eu/Fe][b] (dex) |
|---|---|---|---|---|---|
| 25 Ori | −0.14 ± 0.06 (5) | 2.07 ± 0.02 (2) | 0.02 ± 0.01 (2) | – | – |
| ASCC 50 | −0.12 ± 0.03 (3) | 2.06 ± 0.02 (3) | 0.01 ± 0.02 (3) | – | – |
| Berkeley 21 | −0.18 ± 0.02 (51) | 2.02 ± 0.05 (40) | 0.00 ± 0.07 (40) | 0.50 ± 0.02 (4) | 0.15 ± 0.02 (4) |
| Berkeley 22 | −0.50 ± 0.03 (48) | 1.98 ± 0.03 (31) | 0.33 ± 0.03 (31) | – | – |
| Berkeley 30 | −0.38 ± 0.07 (9) | 2.21 ± 0.04 (4) | 0.34 ± 0.11 (4) | – | – |
| Berkeley 31 | −0.50 ± 0.02 (72) | 1.94 ± 0.03 (60) | 0.32 ± 0.02 (60) | – | – |
| Berkeley 32 | −0.39 ± 0.01 (189) | 2.01 ± 0.01 (160) | 0.25 ± 0.02 (160) | 0.49 ± 0.01 (4) | 0.21 ± 0.01 (4) |
| Berkeley 36 | −0.27 ± 0.01 (151) | 2.12 ± 0.03 (125) | 0.25 ± 0.02 (124) | 0.52 ± 0.05 (2) | 0.21 ± 0.01 (2) |
| Berkeley 39 | −0.24 ± 0.01 (433) | 1.99 ± 0.01 (393) | 0.11 ± 0.01 (393) | 0.53 ± 0.01 (9) | 0.10 ± 0.01 (9) |
| Berkeley 44 | 0.19 ± 0.01 (17) | 2.47 ± 0.06 (17) | 0.11 ± 0.06 (17) | – | – |
| Berkeley 73 | −0.41 ± 0.05 (17) | 2.14 ± 0.05 (11) | 0.37 ± 0.05 (11) | – | – |
| Berkeley 75 | −0.73 ± 0.02 (24) | 1.96 ± 0.03 (26) | 0.56 ± 0.03 (24) | – | – |
| Berkeley 81 | 0.00 ± 0.04 (10) | 2.24 ± 0.07 (6) | 0.00 ± 0.07 (6) | – | – |
| Blanco 1 | −0.06 ± 0.01 (4) | 2.00 ± 0.02 (4) | −0.10 ± 0.01 (4) | – | – |
| Collinder 69 | −0.14 ± 0.06 (6) | 2.20 ± 0.02 (4) | 0.13 ± 0.01 (4) | – | – |
| Collinder 197 | −0.08 ± 0.11 (6) | 2.08 ± 0.10 (4) | 0.08 ± 0.08 (4) | – | – |
| Czernik 24 | −0.20 ± 0.02 (30) | 1.98 ± 0.03 (25) | 0.06 ± 0.04 (25) | 0.49 ± 0.01 (2) | 0.12 ± 0.01 (2) |
| Czernik 30 | −0.58 ± 0.02 (49) | 2.02 ± 0.02 (40) | 0.47 ± 0.02 (40) | – | – |
| ESO 92-05 | −0.43 ± 0.04 (18) | 1.64 ± 0.07 (10) | −0.08 ± 0.07 (10) | – | – |
| $\gamma$ Vel | −0.13 ± 0.01 (2) | 2.19 ± 0.02 (2) | 0.15 ± 0.02 (2) | – | – |
| Haffner 10 | −0.22 ± 0.01 (179) | 2.34 ± 0.02 (149) | 0.40 ± 0.02 (149) | 0.70 ± 0.02 (4) | 0.29 ± 0.03 (4) |
| IC 2391 | −0.20 ± 0.09 (3) | – | – | – | – |
| IC 2602 | −0.01 ± 0.01 (3) | 2.28 ± 0.02 (2) | 0.12 ± 0.02 (2) | – | – |
| IC 4665 | −0.05 ± 0.02 (9) | 2.22 ± 0.06 (8) | 0.13 ± 0.05 (8) | – | – |
| NGC 2141 | −0.11 ± 0.01 (503) | 2.29 ± 0.01 (446) | 0.24 ± 0.01 (446) | – | – |
| NGC 2158 | −0.20 ± 0.01 (219) | 2.23 ± 0.02 (166) | 0.27 ± 0.02 (163) | 0.66 ± 0.02 (4) | 0.28 ± 0.02 (4) |
| NGC 2232 | −0.09 ± 0.05 (5) | 2.21 ± 0.08 (4) | 0.13 ± 0.06 (3) | – | – |
| NGC 2243 | −0.59 ± 0.01 (402) | 1.95 ± 0.01 (302) | 0.39 ± 0.01 (301) | – | – |
| NGC 2244 | −0.12 ± 0.05 (6) | 2.47 ± 0.08 (7) | 0.46 ± 0.07 (5) | – | – |
| NGC 2264 | −0.30 ± 0.10 (2) | – | – | – | – |
| NGC 2355 | −0.22 ± 0.01 (80) | 1.83 ± 0.04 (63) | −0.12 ± 0.03 (63) | – | – |
| NGC 2420 | −0.24 ± 0.01 (327) | 2.04 ± 0.01 (313) | 0.12 ± 0.01 (311) | 0.44 ± 0.03 (2) | 0.12 ± 0.04 (2) |
| NGC 2425 | −0.22 ± 0.01 (106) | 2.25 ± 0.02 (97) | 0.31 ± 0.02 (97) | 0.56 ± 0.02 (3) | 0.18 ± 0.04 (3) |
| NGC 2451a | −0.07 ± 0.02 (6) | 2.17 ± 0.03 (6) | 0.09 ± 0.02 (6) | – | – |
| NGC 2516 | −0.06 ± 0.01 (127) | 2.26 ± 0.01 (127) | 0.17 ± 0.01 (125) | – | – |

– continued on next page

Table E.1: continued.

| Cluster | [Fe/H] (dex) | $A$(Ba)[a] (dex) | [Ba/Fe][b] (dex) | $A$(Eu)[a] (dex) | [Eu/Fe][b] (dex) |
|---|---|---|---|---|---|
| NGC 2547 | −0.06 ± 0.01 (13) | 2.24 ± 0.03 (13) | 0.13 ± 0.03 (13) | – | – |
| NGC 3532 | −0.01 ± 0.01 (174) | 2.18 ± 0.01 (179) | 0.04 ± 0.01 (173) | – | – |
| NGC 4815 | −0.05 ± 0.02 (17) | 2.43 ± 0.07 (16) | 0.33 ± 0.07 (16) | – | – |
| NGC 6005 | 0.05 ± 0.03 (32) | 2.23 ± 0.04 (30) | 0.05 ± 0.03 (29) | – | – |
| NGC 6067 | −0.01 ± 0.03 (22) | 2.35 ± 0.03 (14) | 0.12 ± 0.03 (14) | – | – |
| NGC 6259 | 0.24 ± 0.03 (12) | 2.43 ± 0.06 (8) | 0.13 ± 0.03 (8) | – | – |
| NGC 6281 | 0.00 ± 0.01 (8) | 1.99 ± 0.03 (8) | −0.17 ± 0.02 (8) | – | – |
| NGC 6405 | 0.04 ± 0.01 (20) | 2.22 ± 0.03 (20) | 0.02 ± 0.03 (20) | – | – |
| NGC 6530 | −0.27 ± 0.07 (12) | 1.94 ± 0.06 (7) | 0.13 ± 0.12 (6) | – | – |
| NGC 6633 | −0.04 ± 0.01 (5) | 2.23 ± 0.09 (5) | 0.11 ± 0.09 (5) | – | – |
| NGC 6649 | −0.01 ± 0.01 (3) | – | – | – | – |
| NGC 6705 | 0.04 ± 0.01 (67) | 2.42 ± 0.03 (49) | 0.20 ± 0.03 (49) | – | – |
| NGC 6709 | −0.03 ± 0.01 (34) | 2.32 ± 0.03 (34) | 0.19 ± 0.03 (34) | – | – |
| NGC 6802 | −0.03 ± 0.03 (14) | 2.66 ± 0.07 (12) | 0.52 ± 0.09 (12) | – | – |
| Pismis 15 | −0.19 ± 0.02 (23) | 2.43 ± 0.05 (23) | 0.46 ± 0.05 (23) | – | – |
| Ruprecht 134 | 0.32 ± 0.02 (32) | 2.24 ± 0.06 (27) | −0.26 ± 0.07 (27) | – | – |
| Trumpler 14 | −0.23 ± 0.03 (4) | 2.02 ± 0.05 (4) | 0.08 ± 0.06 (4) | – | – |
| Trumpler 20 | −0.05 ± 0.01 (71) | 2.43 ± 0.03 (72) | 0.31 ± 0.02 (71) | – | – |
| Trumpler 23 | 0.11 ± 0.06 (8) | 2.67 ± 0.07 (8) | 0.37 ± 0.03 (8) | – | – |
| Trumpler 5 | −0.37 ± 0.01 (611) | 2.31 ± 0.01 (505) | 0.51 ± 0.01 (505) | 0.63 ± 0.01 (12) | 0.42 ± 0.02 (12) |

**Notes.**
– Numbers in parentheses indicate the number of stars used for each reported quantity after applying the corresponding quality selection criteria. All analysed stars are cluster members satisfying `MEMQG` ≥ `0.99` from Jackson et al. (2022).
– [Fe/H] is computed using stars satisfying `flag_atm = 0`.
– [a] $A$(Ba) and $A$(Eu) are computed using stars satisfying element-specific quality cuts: `flag_Ba = 0` and `3-sigma_Ba = 0` for Ba, and `flag_Eu = 0` and `3-sigma_Eu = 0` for Eu.
– [b] [Ba/Fe] and [Eu/Fe] are derived using the same criteria as note [a], with the additional requirement `flag_atm = 0`.